\documentclass[a4paper,11pt]{article}

\makeatletter
\gdef\@fpheader{   }
\gdef\@journal{jhep}

\RequirePackage{amsmath}
\RequirePackage{amssymb}
\RequirePackage{epsfig}
\RequirePackage{graphicx}
\RequirePackage[numbers,sort&compress]{natbib}
\RequirePackage{color}
\RequirePackage[
,colorlinks=true
,urlcolor=blue
,anchorcolor=blue
,citecolor=blue
,filecolor=blue
,linkcolor=blue
,menucolor=blue
,pagecolor=blue
,linktocpage=true
,pdfproducer=medialab
,pdfa=true
]{hyperref}

\newif\ifnotoc\notocfalse
\newif\ifemailadd\emailaddfalse
\newif\iftoccontinuous\toccontinuousfalse

\def\@subheader{\@empty}
\def\@keywords{\@empty}
\def\@abstract{\@empty}
\def\@xtum{\@empty}
\def\@dedicated{\@empty}
\def\@arxivnumber{\@empty}
\def\@collaboration{\@empty}
\def\@collaborationImg{\@empty}
\def\@proceeding{\@empty}
\def\@preprint{\@empty}

\newcommand{\subheader}[1]{\gdef\@subheader{#1}}
\newcommand{\keywords}[1]{\if!\@keywords!\gdef\@keywords{#1}\else%
\PackageWarningNoLine{\jname}{Keywords already defined.\MessageBreak Ignoring last definition.}\fi}
\renewcommand{\abstract}[1]{\gdef\@abstract{#1}}
\newcommand{\dedicated}[1]{\gdef\@dedicated{#1}}
\newcommand{\arxivnumber}[1]{\gdef\@arxivnumber{#1}}
\newcommand{\proceeding}[1]{\gdef\@proceeding{#1}}
\newcommand{\xtumfont}[1]{\textsc{#1}}
\newcommand{\correctionref}[3]{\gdef\@xtum{\xtumfont{#1} \href{#2}{#3}}}
\newcommand\jname{JHEP}
\newcommand\acknowledgments{\section*{Acknowledgments}}

\newcommand\preprint[1]{\gdef\@preprint{\hfill #1}}

\newcommand\note[2][]{%
\if!#1!%
\stepcounter{footnote}\footnotetext{#2}%
\else%
{\renewcommand\thefootnote{#1}%
\footnotetext{#2}}%
\fi}

\newtoks\auth@toks
\renewcommand{\author}[2][]{%
  \if!#1!%
    \auth@toks=\expandafter{\the\auth@toks#2\ }%
  \else
    \auth@toks=\expandafter{\the\auth@toks#2$^{#1}$\ }%
  \fi
}

\newtoks\affil@toks\newif\ifaffil\affilfalse
\newcommand{\affiliation}[2][]{%
\affiltrue
  \if!#1!%
    \affil@toks=\expandafter{\the\affil@toks{\item[]#2}}%
  \else
    \affil@toks=\expandafter{\the\affil@toks{\item[$^{#1}$]#2}}%
  \fi
}

\newtoks\email@toks\newcounter{email@counter}%
\newcommand{\emailAdd}[1]{%
\emailaddtrue%
\ifnum\theemail@counter>0\email@toks=\expandafter{\the\email@toks, \@email{#1}}%
\else\email@toks=\expandafter{\the\email@toks\@email{#1}}%
\fi\stepcounter{email@counter}}
\newcommand{\@email}[1]{\href{mailto:#1}{\tt #1}}

\newcommand*\collaboration[1]{\gdef\@collaboration{#1}}
\newcommand*\collaborationImg[2][]{\gdef\@collaborationImg{#2}}

\newcommand\afterLogoSpace{\smallskip}
\newcommand\afterSubheaderSpace{\vskip3pt plus 2pt minus 1pt}
\newcommand\afterProceedingsSpace{\vskip21pt plus0.4fil minus15pt}
\newcommand\afterTitleSpace{\vskip23pt plus0.06fil minus13pt}
\newcommand\afterRuleSpace{\vskip23pt plus0.06fil minus13pt}
\newcommand\afterCollaborationSpace{\vskip3pt plus 2pt minus 1pt}
\newcommand\afterCollaborationImgSpace{\vskip3pt plus 2pt minus 1pt}
\newcommand\afterAuthorSpace{\vskip5pt plus4pt minus4pt}
\newcommand\afterAffiliationSpace{\vskip3pt plus3pt}
\newcommand\afterEmailSpace{\vskip16pt plus9pt minus10pt\filbreak}
\newcommand\afterXtumSpace{\par\bigskip}
\newcommand\afterAbstractSpace{\vskip16pt plus9pt minus13pt}
\newcommand\afterKeywordsSpace{\vskip16pt plus9pt minus13pt}
\newcommand\afterArxivSpace{\vskip3pt plus0.01fil minus10pt}
\newcommand\afterDedicatedSpace{\vskip0pt plus0.01fil}
\newcommand\afterTocSpace{\bigskip\medskip}
\newcommand\afterTocRuleSpace{\bigskip\bigskip}
\newlength{\affiliationsSep}
\newcommand\beforetochook{\pagestyle{myplain}\pagenumbering{roman}}

\DeclareFixedFont\trfont{OT1}{phv}{b}{sc}{11}

\renewcommand\maketitle{
\pagestyle{empty}
\thispagestyle{titlepage}
\setcounter{page}{0}
\noindent{\small\scshape\@fpheader}\@preprint\par
\afterLogoSpace
\if!\@subheader!\else\noindent{\trfont{\@subheader}}\fi
\afterSubheaderSpace
\if!\@proceeding!\else\noindent{\sc\@proceeding}\fi
\afterProceedingsSpace
{\LARGE\flushleft\sffamily\bfseries\@title\par}
\afterTitleSpace
\hrule height 1.5\p@%
\afterRuleSpace
\if!\@collaboration!\else
{\Large\bfseries\sffamily\raggedright\@collaboration}\par
\afterCollaborationSpace
\fi
\if!\@collaborationImg!\else
{\normalsize\bfseries\sffamily\raggedright\@collaborationImg}\par
\afterCollaborationImgSpace
\fi
{\bfseries\raggedright\sffamily\the\auth@toks\par}
\afterAuthorSpace
\ifaffil\begin{list}{}{%
\setlength{\leftmargin}{0.28cm}%
\setlength{\labelsep}{0pt}%
\setlength{\itemsep}{\affiliationsSep}%
\setlength{\topsep}{-\parskip}}
\itshape\small%
\the\affil@toks
\end{list}\fi
\afterAffiliationSpace
\ifemailadd 
\noindent\hspace{0.28cm}\begin{minipage}[l]{.9\textwidth}
\begin{flushleft}
\textit{E-mail:} \the\email@toks
\end{flushleft}
\end{minipage}
\else 
\PackageWarningNoLine{\jname}{E-mails are missing.\MessageBreak Plese use \protect\emailAdd\space macro to provide e-mails.}
\fi
\afterEmailSpace
\if!\@xtum!\else\noindent{\@xtum}\afterXtumSpace\fi
\if!\@abstract!\else\noindent{\renewcommand\baselinestretch{.9}\textsc{Abstract:}}\ \@abstract\afterAbstractSpace\fi
\if!\@keywords!\else\noindent{\textsc{Keywords:}} \@keywords\afterKeywordsSpace\fi
\if!\@arxivnumber!\else\noindent{\textsc{ArXiv ePrint:}} \href{http://arxiv.org/abs/\@arxivnumber}{\@arxivnumber}\afterArxivSpace\fi
\if!\@dedicated!\else\vbox{\small\it\raggedleft\@dedicated}\afterDedicatedSpace\fi
\ifnotoc\else
\iftoccontinuous\else\newpage\fi
\beforetochook\hrule
\tableofcontents
\afterTocSpace
\hrule
\afterTocRuleSpace
\fi
\setcounter{footnote}{0}
\pagestyle{myplain}\pagenumbering{arabic}
} 

\renewcommand{\baselinestretch}{1.1}\normalsize
\divide\@tempdima\baselineskip
\@tempcnta=\@tempdima
\skip\@mpfootins = \skip\footins
\renewcommand{\@dotsep}{10000}

\newcommand\ps@myplain{
\pagenumbering{arabic}
\renewcommand\@oddfoot{\hfill-- \thepage\ --\hfill}
\renewcommand\@oddhead{}}
\let\ps@plain=\ps@myplain

\newcommand\ps@titlepage{\renewcommand\@oddfoot{}\renewcommand\@oddhead{}}

\numberwithin{equation}{section}

\renewcommand\section{\@startsection{section}{1}{\z@}%
                                   {-3.5ex \@plus -1.3ex \@minus -.7ex}%
                                   {2.3ex \@plus.4ex \@minus .4ex}%
                                   {\normalfont\large\bfseries}}
\renewcommand\subsection{\@startsection{subsection}{2}{\z@}%
                                   {-2.3ex\@plus -1ex \@minus -.5ex}%
                                   {1.2ex \@plus .3ex \@minus .3ex}%
                                   {\normalfont\normalsize\bfseries}}
\renewcommand\subsubsection{\@startsection{subsubsection}{3}{\z@}%
                                   {-2.3ex\@plus -1ex \@minus -.5ex}%
                                   {1ex \@plus .2ex \@minus .2ex}%
                                   {\normalfont\normalsize\bfseries}}
\renewcommand\paragraph{\@startsection{paragraph}{4}{\z@}%
                                   {1.75ex \@plus1ex \@minus.2ex}%
                                   {-1em}%
                                   {\normalfont\normalsize\bfseries}}
\renewcommand\subparagraph{\@startsection{subparagraph}{5}{\parindent}%
                                   {1.75ex \@plus1ex \@minus .2ex}%
                                   {-1em}%
                                   {\normalfont\normalsize\bfseries}}

\def\fnum@figure{\textbf{\figurename\nobreakspace\thefigure}}
\def\fnum@table{\textbf{\tablename\nobreakspace\thetable}}

\long\def\@makecaption#1#2{%
  \vskip\abovecaptionskip
  \sbox\@tempboxa{\small #1. #2}%
  \ifdim \wd\@tempboxa >\hsize
    \small #1. #2\par
  \else
    \global \@minipagefalse
    \hb@xt@\hsize{\hfil\box\@tempboxa\hfil}%
  \fi
  \vskip\belowcaptionskip}

\renewenvironment{thebibliography}[1]{%
\begin{oldthebibliography}{#1}%
\small%
\raggedright%
\setlength{\itemsep}{5pt plus 0.2ex minus 0.05ex}%
}%
{%
\end{oldthebibliography}%
}

\makeatother

\usepackage[T1]{fontenc} 
\usepackage{romannum} 

\title{{\boldmath Constructing effective action for gravitational field by effective potential method}}

\author[a]{Shi-Lin Li,}
\author[a]{Yu-Jie Chen,}
\author[b,1]{Yu-Zhu Chen,}\note{chenyuzhu@nankai.edu.cn}
\author[c,1]{Wen-Du Li,}\note{liwendu@tjnu.edu.cn.}
\author[a,3]{and Wu-Sheng Dai}\note{daiwusheng@tju.edu.cn.}

\affiliation[a]{Department of Physics, Tianjin University, Tianjin 300350, P.R. China}
\affiliation[b]{Theoretical Physics Division, Chern Institute of Mathematics, Nankai University, Tianjin, 300071, P. R. China}
\affiliation[c]{College of Physics and Materials Science, Tianjin Normal University, Tianjin 300387, PR China}

\abstract{The aim of this paper is to construct a quantum effective action for
gravitational fields by the effective potential method in quantum field
theory. The minimum of the quantum effective action gives an equation of
quantum fluctuations. We discuss the quantum fluctuation in the flat spacetime
and in the Schwarzschild spacetime. It is shown that a baby spacetime may be
created from a classical vacuum through a quantum fluctuation.
}

\begin{document} 
\maketitle 

\flushbottom

\section{Introduction}

The main aim of this paper is to construct a quantum effective action and an
equation of the quantum fluctuation for gravitational fields by the effective
potential method. The quantum effective action is constructed by the effective
potential method in quantum field theory
\cite{huang1992quarks,weinberg2005quantum}. The spectral method in quantum
field theory also provides a similar treatment
\cite{graham2003negative,graham2003casimir,graham2002calculating,graham1999energy,graham2009spectral,rahi2009scattering,weigel2018spectral}%
. One of the features of this kind of methods is that the classical solution,
preferably exact classical solutions, plays an important role. The classical
solution determines the effective potential in the equation of quantum
fluctuations. This feature is sometimes a disadvantage and sometimes an
advantage. The disadvantage is that the method is difficult to use without an
exact classical solution. Classical field equations are often nonlinear
equations, such as the $\phi^{4}$-field, which is difficult to solve on the
one hand and has infinitely many solutions on the other hand
\cite{li2021duality}. Different classical solutions will give different
effective potentials, and thus giving different equations of quantum
fluctuations, so it is difficult to give a general treatment. However, this
disadvantage is sometimes an advantage in gravity theory. In gravity theory,
years of research have accumulated many important exact solutions of the
Einstein equation \cite{griffiths2009exact,stephani2009exact}. Each solution
of the Einstein equation, the classical solution, leads to an effective
potential for the quantum fluctuation of gravity. In this way, we can
calculate the quantum fluctuation of classical spacetimes given by the
Einstein equation one by one. Although a one-by-one process has its
limitation, it can deal with some important bound-state case, such as the
Schwarzschild spacetime.

This method also has an advantage. The method is based on classical solutions
which can be either scattering states or bound states. The usual
quantum-field-theory method cannot deal with bound states, because it is based
on the scattering perturbation theory. Concretely, in quantum field theory,
one first quantizes free fields, and then adds interactions adiabatically. The
validity of such a treatment is guaranteed by the Gell-mann and Low theorem.
The free field is the simplest scattering-state field, so the quantum-field
method is not suitable to bound states. This is an important difficulty
encountered in quantum field theory at the technical level. It is this
difficulty that limits the application of quantum field theory in, for
example, hadron physics for hadrons are bound states of quarks. The bound
state in hadron physics can only be dealt with by various models, numerical
methods, or the Bethe-Salpeter equation \cite{dai2000electromagnetic}. In
gravity, bound states become more prominent. Many important classical
gravitational solutions are bound states, such as the Schwarzschild spacetime.
This method, though cannot give a general treatment like Feynman diagram
expansions which can only deal with scattering states, can be used to deal
with the bound-state spacetime.

The minimum of the classical action $S$ gives the classical field equation,
\begin{equation}
\delta S=0.
\end{equation}
The minimum of the quantum effective action $\Gamma$ gives the quantum field
equation \cite{huang1992quarks,weinberg2005quantum},
\begin{equation}
\delta\Gamma=0. \label{dGeq0}%
\end{equation}
In this paper, we construct a quantum effective action $\Gamma$\ for gravity
and construct an equation for the quantum fluctuation for gravitational fields.

To calculate a quantum corrected gravitational field $g_{\mu\nu}$, we start
from the classical gravitational field $\bar{g}_{\mu\nu}$, whether it is
strong or weak, bound-state or scattering-state. We write the quantum
corrected metric $g_{\mu\nu}$ as the classical metric $\bar{g}_{\mu\nu}$ which
is the solution of the Einstein equation plus a quantum correction $h_{\mu\nu
}$ in the form of $g_{\mu\nu}=\bar{g}_{\mu\nu}+\hbar h_{\mu\nu}$. In this way,
the information of bound states is embodied in the classical part $\bar
{g}_{\mu\nu}$ and the information of quantum corrections is embodied in
$h_{\mu\nu}$. We will show that the leading-order contribution of the quantum
correction satisfies a linear equation. Of course, if the higher-order quantum
correction is taken into account, the equation of quantum correction will be a
nonlinear equation.

Technically speaking, we construct the quantum effective action of a
gravitational field by the path integral. We first calculate the generating
functional $Z\left[  J\right]  $ which gives the generating functional
$W\left[  J\right]  $. The quantum effective action $\Gamma$\ can be obtained
from $W\left[  J\right]  $ through a Legendre transform. The minimum of the
effective action $\Gamma$, given by Eq. (\ref{dGeq0}), gives the equation of
the quantum corrected field.

The quantum effect on gravity is an important and difficult problem
\cite{woodard2009far,rovelli2011loop,hollands2015quantum}. One fruitful theory
of quantum gravity comes from string theory
\cite{draper2020finite,geng2020distance,bergshoeff2018nonrelativistic,kutasov2015constraining,eberhardt2019string,cano2018alpha,pioline2019string,antoniadis2018effective,cribiori2020sitter,hull2020black,heslop2018m}%
. The quantum gravity is also considered in the frame of loop quantum gravity
\cite{ashtekar1998quantum,ashtekar1999isolated,ashtekar2000quantum,ashtekar2000isolated,ashtekar2001mechanics,ashtekar2002geometry}%
. The quantum effect for gravity, such as the quantum gravity effect in
Oppenheimer-Snyder collapse \cite{kelly2020black}, the creation and
evaporation of black holes
\cite{yang2019quantum,alesci2019quantum,li2018scalar,li2019scattering,li2021scalar}
are considered. Various methods for quantum gravity are developed, such as the
bootstrap theory
\cite{casadio2020bootstrapped,maxfield2019quantum,bautista2019quantum,afkhami2019fast,casadio2020quantum}%
, the renormalization group theory \cite{falls2019aspects}, the conformal
field theory \cite{aprile2018quantum}, and the group field theory formalism
\cite{oriti2018black}. The quantum effect for various gravity models are
considered, including Gauss--Bonnet gravity \cite{hennigar2020lower},
two-dimensional de Sitter spacetime \cite{cotler2020low}, the nonlocal gravity
model \cite{amendola2017quantum}, the JT and CGHS dilaton gravity
\cite{ruzziconi2021conservation}, the Einstein-Gauss-Bonnet-Maxwell gravity
\cite{ma2020bounds,ma2021d,ma2020vacua}, dark energies \cite{perez2019dark}
and universe models \cite{gielen2018cosmological}, and the Minkowskian black
hole and white hole \cite{Chen2021model}. The species problem of the Hawking
radiation is discussed \cite{chen2018entropy}.

In section \ref{QEA}, we solve the quantum effective action for gravitational
fields. In section \ref{EQF}, we consider the equation for quantum
fluctuations by the quantum effective action. In section \ref{flat}, we
consider the quantum fluctuation in the flat spacetime. In section \ref{QFS},
we consider the quantum fluctuation in the Schwarzschild spacetime.
Conclusions are given in section \ref{Conclusion}.

\section{Quantum effective action \label{QEA}}

In this section, we construct the quantum effective action for gravitational fields.

\subsection{Generating functional: $Z\left[  J\right]  $ and $W\left[
J\right]  $}

We first calculate the generating functional $Z\left[  J\right]  $ of
gravitational fields.

We write the quantum corrected metric $g_{\mu\nu}$ as the classical metric
$\bar{g}_{\mu\nu}$ plus a quantum fluctuation $h_{\mu\nu}$:
\begin{equation}
g_{\mu\nu}=\bar{g}_{\mu\nu}+\hbar h_{\mu\nu}. \label{zkdg-0}%
\end{equation}

The classical action of gravitational fields is
\begin{equation}
S=-\int\sqrt{g}Rd^{4}x. \label{jdzyl-0}%
\end{equation}
The minimum of the classical action $S$, given by $\delta S=0$, leads to the
Einstein equation whose solution is the classical gravitational field $\bar
{g}_{\mu\nu}$.

Next we expand the action (\ref{jdzyl-0}) around the classical gravitational
field $\bar{g}_{\mu\nu}$. To do this, we first expand the determinant
$\sqrt{g}$ and the Ricci scalar $R$, respectively.

\paragraph{Expansion of $\sqrt{g}$.}

Expanding $\sqrt{g}$ around the classical gravitational field $\bar{g}_{\mu
\nu}$,%
\begin{align}
\sqrt{g}  &  =\sqrt{\bar{g}}+\int\sqrt{\bar{g}}d^{4}x_{1}\left.  \frac
{\delta\sqrt{g}}{\delta g_{\mu\nu}}\right\vert _{g_{\mu\nu}=\bar{g}_{\mu\nu}%
}\left(  g_{\mu\nu}-\bar{g}_{\mu\nu}\right) \nonumber\\
&  +\frac{1}{2}\int d^{4}x_{1}d^{4}x_{2}\left(  \sqrt{\bar{g}}\right)
^{2}\left.  \frac{\delta^{2}\sqrt{g}}{\delta g_{\mu\nu}\delta g_{\rho\sigma}%
}\right\vert _{g_{\mu\nu}=\bar{g}_{\mu\nu},g_{\rho\sigma}=\bar{g}_{\rho\sigma
}}\left(  g_{\mu\nu}-\bar{g}_{\mu\nu}\right)  \left(  g_{\rho\sigma}-\bar
{g}_{\rho\sigma}\right)  +\cdots, \label{detzk-0}%
\end{align}
substituting Eq. (\ref{zkdg-0}) into Eq. (\ref{detzk-0}), and keeping up to
second order, the harmonic approximation, we arrive at%
\begin{align}
\sqrt{g}  &  =\sqrt{\bar{g}}+\hbar\int\sqrt{\bar{g}}d^{4}x_{1}\left.
\frac{\delta\sqrt{g}}{\delta g_{\mu\nu}}\right\vert _{g_{\mu\nu}=\bar{g}%
_{\mu\nu}}h_{\mu\nu}\nonumber\\
&  +\frac{1}{2}\hbar^{2}\int d^{4}x_{1}d^{4}x_{2}\left(  \sqrt{\bar{g}%
}\right)  ^{2}\left.  \frac{\delta^{2}\sqrt{g}}{\delta g_{\mu\nu}\delta
g_{\rho\sigma}}\right\vert _{g_{\mu\nu}=\bar{g}_{\mu\nu},g_{\rho\sigma}%
=\bar{g}_{\rho\sigma}}h_{\mu\nu}h_{\rho\sigma}+\cdots. \label{detzk-1}%
\end{align}
The variation $\delta\sqrt{g}$ in Eq. (\ref{detzk-1}) is
\cite{weinberg1973gravitation}%
\begin{equation}
\delta\sqrt{g}=\frac{1}{2}\sqrt{g}g^{\mu\nu}\delta g_{\mu\nu}. \label{ghgbf-0}%
\end{equation}
To calculate the functional derivative $\frac{\delta\sqrt{g}}{\delta g_{\mu
\nu}}$, we rewrite Eq. (\ref{ghgbf-0}) as
\cite{burns1991modern,economou2013green}%
\begin{equation}
\delta\sqrt{g}=\int d^{4}x_{1}\delta^{4}\left(  x_{1}-x\right)  \sqrt{g}%
\frac{1}{2}g^{\mu\nu}\delta g_{\mu\nu} \label{detbf-1}%
\end{equation}
and then
\begin{equation}
\frac{\delta\sqrt{g}}{\delta g_{\mu\nu}}=\delta^{4}\left(  x_{1}-x\right)
\frac{1}{2}g^{\mu\nu}. \label{detbf-1-1}%
\end{equation}

Similarly, to calculate the functional derivative $\frac{\delta^{2}\sqrt{g}%
}{\delta g_{\mu\nu}\delta g_{\rho\sigma}}$, we rewrite Eq. (\ref{detbf-1-1})
as \cite{burns1991modern,economou2013green}%
\begin{equation}
\frac{\delta\sqrt{g}}{\delta g_{\mu\nu}}=\int d^{4}x_{2}\delta^{4}\left(
x_{2}-x_{1}\right)  \delta^{4}\left(  x_{1}-x\right)  \frac{1}{2}g^{\mu\nu}.
\label{detbf-2}%
\end{equation}
Then%
\begin{align}
\delta\left(  \frac{\delta\sqrt{g}}{\delta g_{\mu\nu}}\right)   &  =\int
d^{4}x_{2}\delta^{4}\left(  x_{2}-x_{1}\right)  \delta^{4}\left(
x_{1}-x\right)  \frac{1}{2}\delta g^{\mu\nu}\label{detbf-3}\\
&  =-\int d^{4}x_{2}\sqrt{g}\delta^{4}\left(  x_{2}-x_{1}\right)  \delta
^{4}\left(  x_{1}-x\right)  \frac{1}{2}\frac{1}{\sqrt{g}}g^{\mu\rho}%
g^{\nu\sigma}\delta g_{\rho\sigma}, \label{detbf-4}%
\end{align}
where \cite{weinberg1973gravitation}%
\begin{equation}
\delta g^{\mu\nu}=-g^{\mu\rho}g^{\nu\sigma}\delta g_{\rho\sigma} \label{nbf-0}%
\end{equation}
is used. Eq. (\ref{detbf-4}) directly gives functional derivative
\begin{equation}
\frac{\delta^{2}\sqrt{g}}{\delta g_{\mu\nu}\delta g_{\rho\sigma}}=-\delta
^{4}\left(  x_{2}-x_{1}\right)  \delta^{4}\left(  x_{1}-x\right)  \frac{1}%
{2}\frac{1}{\sqrt{g}}g^{\mu\rho}g^{\nu\sigma}\delta g_{\rho\sigma}.
\label{detbf-5}%
\end{equation}

Substituting Eqs. (\ref{detbf-2}) and (\ref{detbf-5}) into Eq. (\ref{detzk-1})
and performing the integral give%
\begin{equation}
\sqrt{g}=\sqrt{\bar{g}}\left(  1+\hbar\frac{1}{2}\bar{g}^{\mu\nu}h_{\mu\nu
}\right)  . \label{detzk-2}%
\end{equation}

\paragraph{Expansion of $g^{\mu\nu}$.}

Now we expand the inverse metric $g^{\mu\nu}$.

Similarly, we expand $g^{\mu\nu}$ around the classical gravitational field
$\bar{g}^{\mu\nu}$ as%
\begin{align}
g^{\mu\nu}  &  =\bar{g}^{\mu\nu}+\hbar\int\sqrt{\bar{g}}d^{4}x_{1}\left.
\frac{\delta g^{\mu\nu}}{\delta g_{\rho\sigma}}\right\vert _{g_{\rho\sigma
}=\bar{g}_{\rho\sigma}}h_{\rho\sigma}\nonumber\\
&  +\frac{1}{2}\hbar^{2}\int\left(  \sqrt{\bar{g}}\right)  ^{2}d^{4}x_{1}%
d^{4}x_{2}\left.  \frac{\delta^{2}g^{\mu\nu}}{\delta g_{\rho\sigma}\delta
g_{\alpha\beta}}\right\vert _{g_{\rho\sigma}=\bar{g}_{\rho\sigma}%
,g_{\alpha\beta}=\bar{g}_{\alpha\beta}}h_{\rho\sigma}h_{\alpha\beta}.
\label{nzk-1}%
\end{align}
To calculate the functional derivative $\frac{\delta g^{\mu\nu}}{\delta
g_{\rho\sigma}}$, we rewrite Eq. (\ref{nbf-0}) as%
\begin{equation}
\delta g^{\mu\nu}=\int d^{4}x_{1}\delta^{4}\left(  x_{1}-x\right)  \left(
-g^{\mu\rho}g^{\nu\sigma}\delta g_{\rho\sigma}\right)  . \label{nbf-1}%
\end{equation}
Then%
\begin{equation}
\frac{\delta g^{\mu\nu}}{\delta g_{\rho\sigma}}=-\delta^{4}\left(
x_{1}-x\right)  \frac{1}{\sqrt{g}}g^{\mu\rho}g^{\nu\sigma}. \label{nbf-2}%
\end{equation}

By Eq. (\ref{nbf-2}), we have%
\begin{equation}
\delta\left(  \frac{\delta g^{\mu\nu}}{\delta g_{\rho\sigma}}\right)
=-\delta^{4}\left(  x_{1}-x\right)  \left[  \delta\left(  \frac{1}{\sqrt{g}%
}\right)  g^{\mu\rho}g^{\nu\sigma}+\frac{1}{\sqrt{g}}\delta\left(  g^{\mu\rho
}\right)  g^{\nu\sigma}+\frac{1}{\sqrt{g}}g^{\mu\rho}\delta\left(
g^{\nu\sigma}\right)  \right]  . \label{nbf-3}%
\end{equation}
Substituting Eqs. (\ref{ghgbf-0}) and (\ref{nbf-0}) into Eq. (\ref{nbf-3})
gives%
\begin{align}
\delta\left(  \frac{\delta g^{\mu\nu}}{\delta g_{\rho\sigma}}\right)   &
=\delta^{4}\left(  x_{1}-x\right)  \frac{1}{\sqrt{g}}\left(  \frac{1}{2}%
g^{\mu\rho}g^{\nu\sigma}g^{\alpha\beta}+g^{\mu\alpha}g^{\rho\beta}g^{\nu
\sigma}+g^{\mu\rho}g^{\nu\alpha}g^{\sigma\beta}\right)  \delta g_{\alpha\beta
}\nonumber\\
&  =\int d^{4}x_{2}\delta^{4}\left(  x_{2}-x_{1}\right)  \left[  \delta
^{4}\left(  x_{1}-x\right)  \right. \nonumber\\
&  \left.  \times\frac{1}{\sqrt{g}}\left(  \frac{1}{2}g^{\mu\rho}g^{\nu\sigma
}g^{\alpha\beta}+g^{\mu\alpha}g^{\rho\beta}g^{\nu\sigma}+g^{\mu\rho}%
g^{\nu\alpha}g^{\sigma\beta}\right)  \delta g_{\alpha\beta}\right]  ,
\label{nbf-4}%
\end{align}
so%
\begin{align}
\frac{\delta^{2}g^{\mu\nu}}{\delta g_{\alpha\beta}\delta g_{\rho\sigma}}  &
=\delta^{4}\left(  x_{2}-x_{1}\right)  \delta^{4}\left(  x_{1}-x\right)
\frac{1}{\left(  \sqrt{g}\right)  ^{2}}\nonumber\\
&  \times\left(  \frac{1}{2}g^{\mu\rho}g^{\nu\sigma}g^{\alpha\beta}%
+g^{\mu\alpha}g^{\rho\beta}g^{\nu\sigma}+g^{\mu\rho}g^{\nu\alpha}%
g^{\sigma\beta}\right)  . \label{nbf-5}%
\end{align}

Substituting Eqs. (\ref{nbf-2}) and (\ref{nbf-5}) into Eq. (\ref{nzk-1}) and
performing the integral give%
\begin{align}
g^{\mu\nu}  &  =\bar{g}^{\mu\nu}-\hbar\bar{g}^{\mu\rho}\bar{g}^{\nu\sigma
}h_{\rho\sigma}\nonumber\\
&  +\hbar^{2}\frac{1}{2}\left(  \frac{1}{2}\bar{g}^{\mu\rho}\bar{g}^{\nu
\sigma}\bar{g}^{\alpha\beta}+\bar{g}^{\mu\alpha}\bar{g}^{\rho\beta}\bar
{g}^{\nu\sigma}+\bar{g}^{\mu\rho}\bar{g}^{\nu\alpha}\bar{g}^{\sigma\beta
}\right)  h_{\rho\sigma}h_{\alpha\beta}. \label{nzk-2}%
\end{align}

\paragraph{Expansion of $S$.}

We rewrite the action (\ref{jdzyl-0}) as%
\begin{equation}
S=-\int d^{4}x\sqrt{g}g^{\mu\nu}R_{\mu\nu}, \label{jdzyl-1}%
\end{equation}
where the Ricci tensor
\begin{equation}
R_{\mu\nu}=\frac{\partial\Gamma_{\mu\lambda}^{\lambda}}{\partial x^{\nu}%
}-\frac{\partial\Gamma_{\mu\nu}^{\lambda}}{\partial x^{\lambda}}+\Gamma
_{\mu\lambda}^{\eta}\Gamma_{\eta\nu}^{\lambda}-\Gamma_{\mu\nu}^{\eta}%
\Gamma_{\lambda\eta}^{\lambda} \label{Ricci-0}%
\end{equation}
with the connection%
\begin{equation}
\Gamma_{\mu\nu}^{\lambda}=\frac{1}{2}g^{\lambda\kappa}\left(  \frac{\partial
g_{\kappa\nu}}{\partial x^{\mu}}+\frac{\partial g_{\mu\kappa}}{\partial
x^{\nu}}-\frac{\partial g_{\mu\nu}}{\partial x^{\kappa}}\right)  .
\label{Gamma-0}%
\end{equation}

First we expand the connection $\Gamma_{\mu\nu}^{\lambda}$ up to second order:%
\begin{align}
\Gamma_{\mu\nu}^{\lambda}  &  =\bar{\Gamma}_{\mu\nu}^{\lambda}+\hbar\frac
{1}{2}\bar{g}^{\lambda\kappa}\left(  \bar{D}_{\mu}h_{\kappa\nu}+\bar{D}_{\nu
}h_{\mu\kappa}-\bar{D}_{\kappa}h_{\mu\nu}\right) \nonumber\\
&  -\hbar^{2}\left[  \frac{1}{2}\bar{g}^{\lambda\rho}\bar{g}^{\kappa\sigma
}h_{\rho\sigma}\left(  \bar{D}_{\mu}h_{\kappa\nu}+\bar{D}_{\nu}h_{\mu\kappa
}-\bar{D}_{\kappa}h_{\mu\nu}\right)  \right. \nonumber\\
&  \left.  -\frac{1}{2}\left(  \frac{1}{2}\bar{g}^{\lambda\rho}\bar{g}%
^{\alpha\beta}\Gamma_{\mu\nu}^{\sigma}+\bar{g}^{\lambda\alpha}\bar{g}%
^{\rho\beta}\Gamma_{\mu\nu}^{\sigma}-\bar{g}^{\lambda\rho}\bar{g}^{\beta
\sigma}\Gamma_{\mu\nu}^{\alpha}\right)  h_{\rho\sigma}h_{\alpha\beta}\right]
, \label{Gamma-00}%
\end{align}
where $\bar{\Gamma}_{\mu\nu}^{\lambda}$ is the connection and $\bar{D}_{\mu}$
is the covariant derivative with the classical metric $\bar{g}^{\mu\nu}$. For
convenience we\ denote
\begin{align}
P_{\mu\kappa\nu}  &  =\bar{D}_{\mu}h_{\kappa\nu}+\bar{D}_{\nu}h_{\mu\kappa
}-\bar{D}_{\kappa}h_{\mu\nu},\label{P-1}\\
Q_{\mu\nu}^{\lambda\rho\alpha\beta\sigma}  &  =\frac{1}{2}\bar{g}^{\lambda
\rho}\bar{g}^{\alpha\beta}\Gamma_{\mu\nu}^{\sigma}+\bar{g}^{\lambda\alpha}%
\bar{g}^{\rho\beta}\Gamma_{\mu\nu}^{\sigma}-\bar{g}^{\lambda\rho}\bar
{g}^{\beta\sigma}\Gamma_{\mu\nu}^{\alpha}, \label{Q-1}%
\end{align}
and then Eq. (\ref{Gamma-00}) becomes%
\begin{equation}
\Gamma_{\mu\nu}^{\lambda}=\bar{\Gamma}_{\mu\nu}^{\lambda}+\hbar\frac{1}{2}%
\bar{g}^{\lambda\kappa}P_{\mu\kappa\nu}-\hbar^{2}\left(  \frac{1}{2}\bar
{g}^{\lambda\rho}\bar{g}^{\kappa\sigma}h_{\rho\sigma}P_{\mu\kappa\nu}-\frac
{1}{2}Q_{\mu\nu}^{\lambda\rho\alpha\beta\sigma}h_{\rho\sigma}h_{\alpha\beta
}\right)  . \label{Gamma-01}%
\end{equation}
Note that $h_{\mu\nu}=h_{\nu\mu}$.

The expansion of the Ricci tensor, by substituting Eq. (\ref{Gamma-01}) into
(\ref{Ricci-0}), reads%

\begin{align}
R_{\mu\nu}  &  =\bar{R}_{\mu\nu}+\hbar\frac{1}{2}\left[  \bar{D}_{\nu}\left(
\bar{g}^{\lambda\kappa}P_{\mu\kappa\lambda}\right)  -\bar{D}_{\lambda}\left(
\bar{g}^{\lambda\kappa}P_{\mu\kappa\nu}\right)  \right]  +\hbar^{2}\left\{
\frac{1}{2}\left[  \bar{D}_{\nu}\left(  Q_{\mu\lambda}^{\lambda\rho\alpha
\beta\sigma}h_{\rho\sigma}h_{\alpha\beta}-\bar{g}^{\lambda\rho}\bar{g}%
^{\kappa\sigma}P_{\mu\kappa\lambda}h_{\rho\sigma}\right)  \right.  \right.
\nonumber\\
&  \left.  -\bar{D}_{\lambda}\left(  Q_{\mu\nu}^{\lambda\rho\alpha\beta\sigma
}h_{\rho\sigma}h_{\alpha\beta}-\bar{g}^{\lambda\rho}\bar{g}^{\kappa\sigma
}P_{\mu\kappa\nu}h_{\rho\sigma}\right)  \right]  \left.  +\frac{1}{4}\bar
{g}^{\eta\kappa}\bar{g}^{\lambda\chi}\left(  P_{\mu\kappa\lambda}P_{\eta
\chi\nu}-P_{\mu\kappa\nu}P_{\lambda\chi\eta}\right)  \right\}  .
\label{Ricci-03}%
\end{align}

The expansion of the Ricci scalar, by substituting Eqs. (\ref{Ricci-03}) and
(\ref{nzk-2}) into
\begin{equation}
R=g^{\mu\nu}R_{\mu\nu}, \label{Rscalar-0}%
\end{equation}
reads%
\begin{align}
R  &  =\bar{R}+\hbar\left\{  \frac{1}{2}\bar{g}^{\mu\nu}\left[  \bar{D}_{\nu
}\left(  \bar{g}^{\lambda\kappa}P_{\mu\kappa\lambda}\right)  -\bar{D}%
_{\lambda}\left(  \bar{g}^{\lambda\kappa}P_{\mu\kappa\nu}\right)  \right]
-\bar{g}^{\mu\rho}\bar{g}^{\nu\sigma}h_{\rho\sigma}\bar{R}_{\mu\nu}\right\}
\nonumber\\
&  +\hbar^{2}\left\{  -\frac{1}{2}\bar{g}^{\mu\rho}\bar{g}^{\nu\sigma}%
h_{\rho\sigma}\left[  \bar{D}_{\nu}\left(  \bar{g}^{\lambda\kappa}P_{\mu
\kappa\lambda}\right)  -\bar{D}_{\lambda}\left(  \bar{g}^{\lambda\kappa}%
P_{\mu\kappa\nu}\right)  \right]  \right. \nonumber\\
&  +\frac{1}{2}\left(  \frac{1}{2}\bar{g}^{\mu\rho}\bar{g}^{\nu\sigma}\bar
{g}^{\alpha\beta}+\bar{g}^{\mu\alpha}\bar{g}^{\rho\beta}\bar{g}^{\nu\sigma
}+\bar{g}^{\mu\rho}\bar{g}^{\nu\alpha}\bar{g}^{\sigma\beta}\right)
h_{\rho\sigma}h_{\alpha\beta}\bar{R}_{\mu\nu}\nonumber\\
&  +\frac{1}{2}\bar{g}^{\mu\nu}\left[  \bar{D}_{\nu}\left(  -\bar{g}%
^{\lambda\rho}\bar{g}^{\kappa\sigma}h_{\rho\sigma}P_{\mu\kappa\lambda}%
+Q_{\mu\lambda}^{\lambda\rho\alpha\beta\sigma}h_{\rho\sigma}h_{\alpha\beta
}\right)  \right. \nonumber\\
&  \left.  -\bar{D}_{\lambda}\left(  -\bar{g}^{\lambda\rho}\bar{g}%
^{\kappa\sigma}h_{\rho\sigma}P_{\mu\kappa\nu}+Q_{\mu\nu}^{\lambda\rho
\alpha\beta\sigma}h_{\rho\sigma}h_{\alpha\beta}\right)  \right] \nonumber\\
&  \left.  +\frac{1}{4}\bar{g}^{\mu\nu}\left(  \bar{g}^{\eta\kappa}\bar
{g}^{\lambda\chi}P_{\mu\kappa\lambda}P_{\eta\chi\nu}-\bar{g}^{\eta\kappa}%
\bar{g}^{\lambda\chi}P_{\mu\kappa\nu}P_{\lambda\chi\eta}\right)  \right\}  .
\label{Rscalar-1}%
\end{align}

The classical gravitational field $\bar{g}^{\mu\nu}$ satisfies the Einstein
equation,%
\begin{equation}
\bar{R}_{\mu\nu}=0. \label{zkjxz-0}%
\end{equation}
Substituting Eqs. (\ref{zkjxz-0}), (\ref{Rscalar-1}), and (\ref{detzk-2}) into
Eq. (\ref{jdzyl-0}), using \cite{weinberg1973gravitation}
\begin{equation}
\bar{D}_{\nu}V^{\nu}=\frac{1}{\sqrt{\bar{g}}}\partial_{\nu}\sqrt{\bar{g}%
}V^{\nu}, \label{DV}%
\end{equation}
and vanishing the surface term, we arrive at the expansion of the action%
\begin{equation}
S=\frac{\hbar^{2}}{4}\int d^{4}x\sqrt{\bar{g}}h_{\rho\sigma}M^{\rho\sigma
\mu\nu}h_{\mu\nu}, \label{zylejzk-3}%
\end{equation}
where%
\begin{equation}
M^{\rho\sigma\mu\nu}=\left(  \bar{g}^{\mu\nu}\bar{g}^{\rho\sigma}\bar
{g}^{\kappa\lambda}-\frac{1}{2}\bar{g}^{\mu\nu}\bar{g}^{\kappa\rho}\bar
{g}^{\lambda\sigma}+\bar{g}^{\mu\rho}\bar{g}^{\kappa\sigma}\bar{g}^{\lambda
\nu}-\frac{1}{2}\bar{g}^{\mu\rho}\bar{g}^{\lambda\kappa}\bar{g}^{\nu\sigma
}\right)  \bar{D}_{\lambda}\bar{D}_{\kappa} \label{dynb-0-3}%
\end{equation}
or%
\begin{equation}
M^{\rho\sigma\mu\nu}=\bar{g}^{\mu\nu}\left(  \bar{g}^{\rho\sigma}\bar{D}%
^{2}-\frac{1}{2}\bar{D}^{\sigma}\bar{D}^{\rho}\right)  -\bar{g}^{\mu\rho
}\left(  \frac{1}{2}\bar{g}^{\nu\sigma}\bar{D}^{2}-\bar{D}^{\nu}\bar
{D}^{\sigma}\right)  .
\end{equation}

\paragraph{Generating functional: $Z\left[  J\right]  $ and $W\left[
J\right]  $.}

Now we calculate the generating functional for gravitational fields.

We consider an action with a source $J$,%
\begin{equation}
S_{J}=\int\sqrt{g}d^{4}xR+\int\sqrt{g}d^{4}xJ. \label{ohzyl-0-3}%
\end{equation}
The generating functional is%
\begin{equation}
Z\left[  J\right]  =\int\mathcal{D}g_{\mu\nu}\exp\left(  \frac{i}{\hbar
}S\left[  g_{\mu\nu}\right]  +\frac{i}{\hbar}\int\sqrt{g}d^{4}xJ\right)  .
\label{ljjf-0-3}%
\end{equation}
Substituting the expansions (\ref{zkdg-0}), (\ref{detzk-2}), and
(\ref{zylejzk-3}) into the generating functional (\ref{ljjf-0-3}), we have
\begin{equation}
Z\left[  J\right]  =\int\mathcal{D}h_{\mu\nu}\exp\left(  \frac{i\hbar}{4}%
\int\sqrt{\bar{g}}d^{4}xh_{\rho\sigma}M^{\rho\sigma\mu\nu}h_{\mu\nu}\right.
\left.  +\frac{i}{2}\int\sqrt{\bar{g}}d^{4}x\bar{g}^{\mu\nu}h_{\mu\nu}%
J+\frac{i}{\hbar}\int d^{4}x\sqrt{\bar{g}}J\right)  . \label{ljjf-3-3}%
\end{equation}

We formally rewrite the generating functional (\ref{ljjf-3-3}) as%
\begin{equation}
Z\left[  J\right]  =\int\mathcal{D}h\exp\left[  i\hbar\frac{1}{4}\left(
h,Mh\right)  +i\frac{1}{2}\left(  J\bar{g},h\right)  +\frac{i}{\hbar}\int
d^{4}x\sqrt{\bar{g}}J\right]  \label{ZJhM}%
\end{equation}
for convenience.

To perform the path integral, we use the Gaussian integral formula
\cite{ryder1996quantum}%
\begin{equation}
\int\mathcal{D}\phi\exp\left\{  -\left[  \frac{1}{2}\left(  \phi,A\phi\right)
+\left(  b,\phi\right)  +c\right]  \right\}  =\left(  \det A\right)
^{-1/2}\exp\left[  \frac{1}{2}\left(  b,A^{-1}b\right)  -c\right]
.\label{gausse-0-3}%
\end{equation}
By Eq. (\ref{gausse-0-3}), the generating functional (\ref{ZJhM}) becomes%
\begin{equation}
Z\left[  J\right]  =\left[  \det\left(  -\frac{i\hbar}{2}M\right)  \right]
^{-1/2}\exp\left[  -i\frac{1}{4\hbar}\left(  J\bar{g},M^{-1}J\bar{g}\right)
+\frac{i}{\hbar}\int d^{4}x\sqrt{\bar{g}}J\right]  ,\label{scfh-2}%
\end{equation}
or equivalently,%
\begin{align}
Z\left[  J\right]   &  =\left[  \det\left(  -\frac{i\hbar}{2}M\right)
\right]  ^{-1/2}\nonumber\\
&  \times\exp\left\{  \frac{1}{4i\hbar}\int\sqrt{\bar{g}}d^{4}x\sqrt{\bar{g}%
}d^{4}x^{\prime}\bar{g}^{\mu\nu}\left(  \mathbf{x}\right)  J\left(
\mathbf{x}\right)  \left(  M^{-1}\right)  _{\mu\nu\rho\sigma}\left(
\mathbf{x,x}^{\prime}\right)  \bar{g}^{\rho\sigma}\left(  \mathbf{x}^{\prime
}\right)  J\left(  \mathbf{x}^{\prime}\right)  \right.  \nonumber\\
&  \left.  +\frac{i}{\hbar}\int d^{4}x\sqrt{\bar{g}}J\right\}
.\label{ljjf-4-3}%
\end{align}

The generating functional $W\left[  J\right]  $ is defined as%
\begin{equation}
Z\left[  J\right]  =e^{\frac{i}{\hbar}W\left[  J\right]  }. \label{ZW-0-3}%
\end{equation}
From $Z\left[  J\right]  $ given by Eq. (\ref{ljjf-4-3})\ we obtain%
\begin{align}
W\left[  J\right]   &  =-i\hbar\ln Z\left[  J\right] \nonumber\\
&  =\frac{i}{2}\hbar\operatorname{tr}\ln\left(  -i\hbar\frac{1}{2}M\right)
+\int d^{4}x\sqrt{\bar{g}}J.\nonumber\\
&  -\frac{1}{4}\int\sqrt{\bar{g}}d^{4}x\sqrt{\bar{g}}d^{4}x^{\prime}\bar
{g}^{\mu\nu}\left(  \mathbf{x}\right)  J\left(  \mathbf{x}\right)  \left(
M^{-1}\right)  _{\mu\nu\rho\sigma}\left(  \mathbf{x,x}^{\prime}\right)
\bar{g}^{\rho\sigma}\left(  \mathbf{x}^{\prime}\right)  J\left(
\mathbf{x}^{\prime}\right)  . \label{ZW-1-3}%
\end{align}

\subsection{Quantum effective action: Legendre transform}

The quantum effective action is the Legendre transform of the generating
functional $W\left[  J\right]  $:%
\begin{equation}
\Gamma\left[  h_{\mu\nu}\right]  =W\left[  J\right]  -\int d^{4}x\sqrt{\bar
{g}}h_{\mu\nu}\bar{g}^{\mu\nu}J \label{yxzyl-1-3}%
\end{equation}
with
\begin{equation}
h_{\mu\nu}=\frac{\delta W\left[  J\right]  }{\delta\left(  \bar{g}^{\mu\nu
}J\right)  }. \label{lrd-0-3}%
\end{equation}
Here we rewrite $\Gamma\left[  g_{\mu\nu}\right]  =\Gamma\left[  \bar{g}%
_{\mu\nu}+h_{\mu\nu}\right]  $ $=\Gamma\left[  h_{\mu\nu}\right]  $ for the
classical solution $\bar{g}_{\mu\nu}$ is a known function in the functional
$\Gamma$.

Rewrite the generating functional $W\left[  J\right]  $ given by Eq.
(\ref{ZW-1-3}) as
\begin{align}
W\left[  J\right]   &  =\frac{i\hbar}{2}\operatorname{tr}\ln\left[
-i\hbar\frac{1}{2}M\right]  +\frac{1}{N}\int d^{4}x\sqrt{\bar{g}}\bar{g}%
_{\mu\nu}\bar{g}^{\mu\nu}J\nonumber\\
&  -\frac{1}{4}\int\sqrt{\bar{g}}d^{4}x\sqrt{\bar{g}}d^{4}x^{\prime}J\left(
\mathbf{x}\right)  \bar{g}^{\rho\sigma}\left(  \mathbf{x}\right)  \left(
M^{-1}\right)  _{\rho\sigma\mu\nu}\left(  \mathbf{x},\mathbf{x}^{\prime
}\right)  J\left(  \mathbf{x}^{\prime}\right)  \bar{g}^{\mu\nu}\left(
\mathbf{x}^{\prime}\right)  , \label{ZW-1-3-0}%
\end{align}
where $N$ is the dimension of the spacetime.

Taking variation of $W\left[  J\right]  $ gives
\begin{align}
\delta W\left[  J\right]   &  =-\frac{1}{2}\int\sqrt{\bar{g}}d^{4}x\sqrt
{\bar{g}}d^{4}x^{\prime}\delta\left[  J\left(  \mathbf{x}\right)  \bar
{g}^{\rho\sigma}\left(  \mathbf{x}\right)  \right]  \left(  M^{-1}\right)
_{\rho\sigma\mu\nu}\left(  \mathbf{x},\mathbf{x}^{\prime}\right)  J\left(
\mathbf{x}^{\prime}\right)  \bar{g}^{\mu\nu}\left(  \mathbf{x}^{\prime}\right)
\nonumber\\
&  +\frac{1}{N}\int d^{4}x\sqrt{\bar{g}}\delta\left(  \bar{g}^{\mu\nu
}J\right)  \bar{g}_{\mu\nu}. \label{lrd-0-7}%
\end{align}

Substituting Eq. (\ref{lrd-0-7}) into Eq. (\ref{lrd-0-3}) gives
\begin{equation}
2\left(  \frac{1}{N}\bar{g}_{\mu\nu}-h_{\mu\nu}\right)  =\int\sqrt{\bar{g}%
}d^{4}x^{\prime}\left(  M^{-1}\right)  _{\mu\nu\rho\sigma}\left(
\mathbf{x},\mathbf{x}^{\prime}\right)  J\left(  \mathbf{x}^{\prime}\right)
\bar{g}^{\rho\sigma}\left(  \mathbf{x}^{\prime}\right)  . \label{lrd-0-5}%
\end{equation}

Now we solve $J$. The action with the source $J$, Eq. (\ref{ohzyl-0-3}), by
Eqs. (\ref{zylejzk-3}) and (\ref{detzk-2}), reads%
\begin{equation}
S_{J}=\frac{\hbar^{2}}{4}\int d^{4}x\sqrt{\bar{g}}h_{\rho\sigma}M^{\rho
\sigma\mu\nu}h_{\mu\nu}+\frac{1}{2}\hbar\int\sqrt{\bar{g}}d^{4}xh_{\mu\nu}%
\bar{g}^{\mu\nu}J+\int\sqrt{\bar{g}}d^{4}xJ. \label{jdszk-0}%
\end{equation}

Taking variation of $S_{J}$ and vanishing the surface term give
\begin{align}
\delta S  &  =\frac{\hbar^{2}}{4}\int d^{4}x\sqrt{\bar{g}}\left(  \delta
h_{\mu\nu}\right)  \left[  \left(  \bar{g}^{\rho\sigma}\bar{g}^{\mu\nu}\bar
{g}^{\kappa\lambda}-\frac{1}{2}\bar{g}^{\rho\sigma}\bar{g}^{\kappa\mu}\bar
{g}^{\lambda\nu}+\bar{g}^{\rho\mu}\bar{g}^{\kappa\nu}\bar{g}^{\lambda\sigma
}-\frac{1}{2}\bar{g}^{\rho\mu}\bar{g}^{\lambda\kappa}\bar{g}^{\sigma\nu
}\right)  \bar{D}_{\lambda}\bar{D}_{\kappa}h_{\rho\sigma}\right. \nonumber\\
&  \left.  +\left(  \bar{g}^{\mu\nu}\bar{g}^{\rho\sigma}\bar{g}^{\lambda
\kappa}-\frac{1}{2}\bar{g}^{\mu\nu}\bar{g}^{\lambda\rho}\bar{g}^{\kappa\sigma
}+\bar{g}^{\mu\rho}\bar{g}^{\lambda\sigma}\bar{g}^{\kappa\nu}-\frac{1}{2}%
\bar{g}^{\mu\rho}\bar{g}^{\kappa\lambda}\bar{g}^{\nu\sigma}\right)  \bar
{D}_{\lambda}\bar{D}_{\kappa}h_{\rho\sigma}\right] \nonumber\\
&  +\frac{1}{2}\hbar\int d^{4}x\sqrt{\bar{g}}\left(  \delta h_{\mu\nu}\right)
\bar{g}^{\mu\nu}J+\int d^{4}x\sqrt{\bar{g}}J. \label{jdszk-bf2}%
\end{align}
The minimum of $S_{J}$, given by $\delta S_{J}=0$, gives%
\begin{equation}
\frac{1}{2}\left(  M^{\rho\sigma\mu\nu}+M^{\mu\nu\rho\sigma}\right)  h_{\mu
\nu}=-\bar{g}^{\rho\sigma}J. \label{eq-0}%
\end{equation}

Substituting Eqs. (\ref{lrd-0-5}) and (\ref{eq-0}) into the Legendre transform
(\ref{yxzyl-1-3}) gives the effective action%
\begin{align}
\Gamma\left[  h_{\mu\nu}\right]   &  =\frac{i}{2}\hbar\operatorname{tr}%
\ln\left(  -i\hbar\frac{1}{2}M\right)  +\frac{1}{4}\int d^{4}x\sqrt{\bar{g}%
}h_{\rho\sigma}\left(  M^{\rho\sigma\mu\nu}+M^{\mu\nu\rho\sigma}\right)
h_{\mu\nu}\nonumber\\
&  -\frac{1}{4N}\int d^{4}x\sqrt{\bar{g}}\bar{g}_{\rho\sigma}\left(
M^{\rho\sigma\mu\nu}+M^{\mu\nu\rho\sigma}\right)  h_{\mu\nu}.
\label{dqyxzyl-0-3}%
\end{align}

\section{Equation of quantum fluctuation \label{EQF}}

The minimum of the quantum effective action $\Gamma$ gives the equation of the
quantum fluctuation:%
\begin{equation}
\delta\Gamma=0 \label{zlfc-1-3}%
\end{equation}

Taking variation of the effective action (\ref{dqyxzyl-0-3})%
\begin{align}
\delta\Gamma &  =\frac{1}{4}\int d^{4}x\sqrt{\bar{g}}\left(  \delta
h_{\rho\sigma}\right)  \left(  M^{\rho\sigma\mu\nu}+M^{\mu\nu\rho\sigma
}\right)  h_{\mu\nu}+\frac{1}{4}\int d^{4}x\sqrt{\bar{g}}h_{\rho\sigma}\left(
M^{\rho\sigma\mu\nu}+M^{\mu\nu\rho\sigma}\right)  \left(  \delta h_{\mu\nu
}\right) \nonumber\\
&  -\frac{1}{4N}\int d^{4}x\sqrt{\bar{g}}\bar{g}_{\rho\sigma}\left(
M^{\rho\sigma\mu\nu}+M^{\mu\nu\rho\sigma}\right)  \left(  \delta h_{\mu\nu
}\right)  , \label{dqbf-0-3-1}%
\end{align}
using $\bar{D}_{\lambda}\bar{g}_{\rho\sigma}=0$ and Eq. (\ref{DV}), we have
\begin{align}
\delta\Gamma &  =\frac{1}{4}\int d^{4}x\sqrt{\bar{g}}\left(  \delta
h_{\rho\sigma}\right)  \left(  \bar{g}^{\mu\nu}\bar{g}^{\rho\sigma}\bar
{g}^{\kappa\lambda}-\frac{1}{2}\bar{g}^{\mu\nu}\bar{g}^{\kappa\rho}\bar
{g}^{\lambda\sigma}+\bar{g}^{\mu\rho}\bar{g}^{\kappa\sigma}\bar{g}^{\lambda
\nu}-\frac{1}{2}\bar{g}^{\mu\rho}\bar{g}^{\lambda\kappa}\bar{g}^{\nu\sigma
}\right)  \bar{D}_{\lambda}\bar{D}_{\kappa}h_{\mu\nu}\nonumber\\
&  +\frac{1}{4}\int d^{4}x\sqrt{\bar{g}}\left(  \delta h_{\rho\sigma}\right)
\left(  \bar{g}^{\rho\sigma}\bar{g}^{\mu\nu}\bar{g}^{\kappa\lambda}-\frac
{1}{2}\bar{g}^{\rho\sigma}\bar{g}^{\kappa\mu}\bar{g}^{\lambda\nu}+\bar
{g}^{\rho\mu}\bar{g}^{\kappa\nu}\bar{g}^{\lambda\sigma}-\frac{1}{2}\bar
{g}^{\rho\mu}\bar{g}^{\lambda\kappa}\bar{g}^{\sigma\nu}\right)  \bar
{D}_{\lambda}\bar{D}_{\kappa}h_{\mu\nu}\nonumber\\
&  +\frac{1}{4}\int d^{4}x\sqrt{\bar{g}}\left(  \delta h_{\mu\nu}\right)
\left(  \bar{g}^{\mu\nu}\bar{g}^{\rho\sigma}\bar{g}^{\kappa\lambda}-\frac
{1}{2}\bar{g}^{\mu\nu}\bar{g}^{\kappa\rho}\bar{g}^{\lambda\sigma}+\bar{g}%
^{\mu\rho}\bar{g}^{\kappa\sigma}\bar{g}^{\lambda\nu}-\frac{1}{2}\bar{g}%
^{\mu\rho}\bar{g}^{\lambda\kappa}\bar{g}^{\nu\sigma}\right)  \bar{D}_{\kappa
}\bar{D}_{\lambda}h_{\rho\sigma}\nonumber\\
&  +\frac{1}{4}\int d^{4}x\sqrt{\bar{g}}\left(  \delta h_{\mu\nu}\right)
\left(  \bar{g}^{\rho\sigma}\bar{g}^{\mu\nu}\bar{g}^{\kappa\lambda}-\frac
{1}{2}\bar{g}^{\rho\sigma}\bar{g}^{\kappa\mu}\bar{g}^{\lambda\nu}+\bar
{g}^{\rho\mu}\bar{g}^{\kappa\nu}\bar{g}^{\lambda\sigma}-\frac{1}{2}\bar
{g}^{\rho\mu}\bar{g}^{\lambda\kappa}\bar{g}^{\sigma\nu}\right)  \bar
{D}_{\kappa}\bar{D}_{\lambda}h_{\rho\sigma}\nonumber\\
&  +\frac{1}{4}\int d^{4}x\partial_{\lambda}\left[  \sqrt{\bar{g}}%
h_{\rho\sigma}\left(  \bar{g}^{\mu\nu}\bar{g}^{\rho\sigma}\bar{g}%
^{\kappa\lambda}-\frac{1}{2}\bar{g}^{\mu\nu}\bar{g}^{\kappa\rho}\bar
{g}^{\lambda\sigma}+\bar{g}^{\mu\rho}\bar{g}^{\kappa\sigma}\bar{g}^{\lambda
\nu}-\frac{1}{2}\bar{g}^{\mu\rho}\bar{g}^{\lambda\kappa}\bar{g}^{\nu\sigma
}\right)  \bar{D}_{\kappa}\left(  \delta h_{\mu\nu}\right)  \right]
\nonumber\\
&  -\frac{1}{4}\int d^{4}x\partial_{\kappa}\left[  \sqrt{\bar{g}}\left(
\bar{D}_{\lambda}h_{\rho\sigma}\right)  \left(  \bar{g}^{\mu\nu}\bar{g}%
^{\rho\sigma}\bar{g}^{\kappa\lambda}-\frac{1}{2}\bar{g}^{\mu\nu}\bar
{g}^{\kappa\rho}\bar{g}^{\lambda\sigma}+\bar{g}^{\mu\rho}\bar{g}^{\kappa
\sigma}\bar{g}^{\lambda\nu}-\frac{1}{2}\bar{g}^{\mu\rho}\bar{g}^{\lambda
\kappa}\bar{g}^{\nu\sigma}\right)  \left(  \delta h_{\mu\nu}\right)  \right]
\nonumber\\
&  +\frac{1}{4}\int d^{4}x\partial_{\lambda}\left[  \sqrt{\bar{g}}%
h_{\rho\sigma}\left(  \bar{g}^{\rho\sigma}\bar{g}^{\mu\nu}\bar{g}%
^{\kappa\lambda}-\frac{1}{2}\bar{g}^{\rho\sigma}\bar{g}^{\kappa\mu}\bar
{g}^{\lambda\nu}+\bar{g}^{\rho\mu}\bar{g}^{\kappa\nu}\bar{g}^{\lambda\sigma
}-\frac{1}{2}\bar{g}^{\rho\mu}\bar{g}^{\lambda\kappa}\bar{g}^{\sigma\nu
}\right)  \bar{D}_{\kappa}\left(  \delta h_{\mu\nu}\right)  \right]
\nonumber\\
&  -\frac{1}{4}\int d^{4}x\partial_{\kappa}\left[  \sqrt{\bar{g}}\left(
\bar{D}_{\lambda}h_{\rho\sigma}\right)  \left(  \bar{g}^{\rho\sigma}\bar
{g}^{\mu\nu}\bar{g}^{\kappa\lambda}-\frac{1}{2}\bar{g}^{\rho\sigma}\bar
{g}^{\kappa\mu}\bar{g}^{\lambda\nu}+\bar{g}^{\rho\mu}\bar{g}^{\kappa\nu}%
\bar{g}^{\lambda\sigma}-\frac{1}{2}\bar{g}^{\rho\mu}\bar{g}^{\lambda\kappa
}\bar{g}^{\sigma\nu}\right)  \left(  \delta h_{\mu\nu}\right)  \right]
\nonumber\\
&  -\frac{1}{4N}\int d^{4}x\partial_{\lambda}\left[  \sqrt{\bar{g}}\bar
{g}_{\rho\sigma}\left(  \bar{g}^{\mu\nu}\bar{g}^{\rho\sigma}\bar{g}%
^{\kappa\lambda}-\frac{1}{2}\bar{g}^{\mu\nu}\bar{g}^{\kappa\rho}\bar
{g}^{\lambda\sigma}+\bar{g}^{\mu\rho}\bar{g}^{\kappa\sigma}\bar{g}^{\lambda
\nu}-\frac{1}{2}\bar{g}^{\mu\rho}\bar{g}^{\lambda\kappa}\bar{g}^{\nu\sigma
}\right)  \bar{D}_{\kappa}\left(  \delta h_{\mu\nu}\right)  \right]
\nonumber\\
&  -\frac{1}{4N}\int d^{4}x\partial_{\lambda}\left[  \sqrt{\bar{g}}\bar
{g}_{\rho\sigma}\left(  \bar{g}^{\rho\sigma}\bar{g}^{\mu\nu}\bar{g}%
^{\kappa\lambda}-\frac{1}{2}\bar{g}^{\rho\sigma}\bar{g}^{\kappa\mu}\bar
{g}^{\lambda\nu}\right.  \right. \nonumber\\
&  \left.  \left.  +\bar{g}^{\rho\mu}\bar{g}^{\kappa\nu}\bar{g}^{\lambda
\sigma}-\frac{1}{2}\bar{g}^{\rho\mu}\bar{g}^{\lambda\kappa}\bar{g}^{\sigma\nu
}\right)  \bar{D}_{\kappa}\left(  \delta h_{\mu\nu}\right)  \right]  .
\label{dqbf-00-3-3}%
\end{align}
Vanishing the surface term, we arrive at%
\begin{equation}
\delta\Gamma=\frac{1}{2}\int\sqrt{\bar{g}}d^{4}x\left(  \delta h_{\mu\nu
}\right)  \left(  M^{\rho\sigma\mu\nu}+M^{\mu\nu\rho\sigma}\right)
h_{\rho\sigma}. \label{dqbf-0-3}%
\end{equation}

The equation of quantum fluctuation $h_{\mu\nu}$ by Eq. (\ref{zlfc-1-3}) is
\begin{equation}
\left(  M^{\rho\sigma\mu\nu}+M^{\mu\nu\rho\sigma}\right)  h_{\rho\sigma}=0,
\label{zlfc-2-3}%
\end{equation}
or equivalently,%
\begin{equation}
\left[  \left(  2\bar{g}^{\mu\nu}\bar{g}^{\rho\sigma}-\bar{g}^{\mu\rho}\bar
{g}^{\sigma\nu}\right)  \bar{D}^{2}-\frac{1}{2}\bar{g}^{\mu\nu}\bar{D}^{\rho
}\bar{D}^{\sigma}-\frac{1}{2}\bar{g}^{\rho\sigma}\bar{D}^{\nu}\bar{D}^{\mu
}+\bar{g}^{\mu\rho}\bar{D}^{\sigma}\bar{D}^{\nu}+\bar{g}^{\rho\mu}\bar{D}%
^{\nu}\bar{D}^{\sigma}\right]  h_{\rho\sigma}=0. \label{zlfc-2-4}%
\end{equation}

Eq. (\ref{zlfc-2-3}) is the equation of the leading-order quantum fluctuation.

\section{Quantum fluctuation in flat spacetime \label{flat}}

In this section, we consider the quantum fluctuation in flat spacetime. Flat
space has the largest symmetry, and the symmetry of spacetime after quantum
fluctuation is determined by the symmetry of the quantum fluctuation

Just like in spontaneous magnetization, the system is isotropic before
magnetization. The direction of magnetization is determined by the direction
of external disturbance. In the calculation of magnetization, to simulate the
disturbance, one first add an external magnetic field to the system, and, at
the end of the calculation, take the external magnetic field to zero. That is,
the effect of this external magnetic field is to point a magnetization direction.

For the sake of calculation, we take the coordinate of flat spacetime to match
the symmetry of the fluctuation. In simple cases, for example, for spherically
symmetric spacetime, we use spherical coordinates. In complex cases, we use
the following method to choose the coordinates for describing flat spacetime.
Let us take the Kerr spacetime as an example. If we want the quantum corrected
spacetime to have a similar symmetry to the Kerr spacetime, we can take the
metric of flat spacetime in the following way. It is known that the zero-mass
Kerr spacetime is a flat spacetime. Therefore, we can write the metric of flat
spacetime by first writing down a Kerr metric and then vanishing the mass. The
curvatures described by the Riemann tensor is of course zero. Note that here
in the following case, the Ricci scalar and tensor are zero for the classical
spacetime we considered is the vacuum solution of the Einstein equation.

\subsection{Spherically symmetric quantum fluctuation}

In this section, we consider a flat spacetime with a spherically symmetric
quantum fluctuation.

We represent the line element of the flat spacetime in spherical coordinates,
\begin{equation}
ds^{2}=dt^{2}-dr^{2}-r^{2}d\theta^{2}-r^{2}\sin^{2}\theta d\varphi^{2},
\end{equation}
where the metric is written in spherical coordinates. Supposing the quantum
corrected spacetime is spherically symmetric, we take the line element of the
quantum corrected spacetime as
\begin{equation}
ds^{2}=\left[  1+f\left(  r\right)  \right]  dt^{2}-\left[  1+g\left(
r\right)  \right]  dr^{2}-r^{2}d\theta^{2}-r^{2}\sin^{2}\theta d\varphi^{2}.
\label{qdczl-000}%
\end{equation}

The quantum fluctuation is determined by the equation of fluctuation
(\ref{zlfc-2-4}). Substituting Eq. (\ref{qdczl-000}) into Eq. (\ref{zlfc-2-4})
gives the following equations%
\begin{align}
2rf^{\prime}(r)+r^{2}f^{\prime\prime}(r)+2rg^{\prime}(r)+\frac{3}{2}%
r^{2}g^{\prime\prime}(r)-g(r)  &  =0,\nonumber\\
r\left(  -8f^{\prime}(r)-3rf^{\prime\prime}(r)-8g^{\prime}(r)-4rg^{\prime
\prime}(r)\right)  +2g(r)  &  =0,\nonumber\\
r\left(  7f^{\prime}(r)+4rf^{\prime\prime}(r)+7g^{\prime}(r)+3rg^{\prime
\prime}(r)\right)  +2g(r)  &  =0.
\end{align}
Solving these equations gives
\begin{align}
f\left(  r\right)   &  =-C_{1}r^{2}+C_{2}-C\frac{1}{r},\label{pzzl00-0}\\
g\left(  r\right)   &  =C_{1}r^{2}+C\frac{1}{r}. \label{pzzl11-0}%
\end{align}

Requiring that the quantum fluctuation $h_{\mu\nu}$ vanishes at $r\rightarrow
\infty$, we have $C_{1}=0$ and $C_{2}=0$, i.e.,
\begin{align}
f\left(  r\right)   &  =-\frac{C}{r},\label{pzzl00-1}\\
g\left(  r\right)   &  =\frac{C}{r}. \label{pzzl11-1}%
\end{align}
The quantum corrected result then reads%
\begin{equation}
ds^{2}=\left(  1-\frac{C}{r}\right)  dt^{2}-\left(  1+\frac{C}{r}\right)
dr^{2}-r^{2}d\theta^{2}-r^{2}\sin^{2}\theta d\varphi^{2}. \label{pzxzdg-0}%
\end{equation}

A simple analysis shows that this quantum corrected spacetime has
singularities at $r=0$ and $r=C$, which can be checked by calculating the
Ricci scalar, the Ricci tensor, the Riemann tensor, and the Weyl tensor in the
orthogonal frame. This spacetime also has an event horizon and an infinite
redshift surface at $r=C$.

By comparing with a small-mass Schwarzschild spacetime whose metric is the
leading term of the expansion around $M=0$ of the Schwarzschild metric,%
\begin{equation}
ds^{2}\simeq\left(  1-\frac{2M}{r}\right)  dt^{2}-\left(  1+\frac{2M}%
{r}\right)  dr^{2}-r^{2}d\theta^{2}-r^{2}\sin^{2}\theta d\varphi^{2},
\end{equation}
we can see that $C$ plays a role of mass, i.e., $C=2M$. The constant $C$ comes
from a quantum fluctuation, so the mass $M$ must be very small and should be
of the magnitude of the Planck mass.

That is, if there occurs a spherically symmetric quantum fluctuation in a flat
spacetime, a Schwarzschild spacetime may be created.

It should be emphasized that the quantum corrected flat spacetime is no long a
classical vacuum.

\subsection{Axially symmetric quantum fluctuation}

In this section, we consider a flat spacetime with an axially symmetrical
quantum fluctuation.

For convenience of calculations, we write the metric of a flat space as a
zero-mass Kerr spacetime who is a flat spacetime with zero curvatures, e.g.,
a\ zero Riemann curvature.

More concretely, the Kerr metric is \cite{ohanian2013gravitation}%
\begin{equation}
ds^{2}=dt^{2}-\frac{\rho^{2}}{\Delta}dr^{2}-\rho^{2}d\theta^{2}-\left(
r^{2}+a^{2}\right)  \sin^{2}\theta d\phi^{2}-2\frac{mr}{\rho^{2}}\left(
dt-a\sin^{2}\theta d\phi\right)  ^{2}, \label{kerr-0}%
\end{equation}
where $\rho^{2}=r^{2}+a^{2}\cos^{2}\theta$ and $\Delta=r^{2}-2mr+a^{2}$ with
$a$ the angular momentum per unit mass. By the coordinate transformation
$\cos\theta=x$, we arrive at the Kerr metric represented by the coordinate
$\left(  t,r,x,\varphi\right)  $:%
\begin{align}
ds^{2}  &  =\left[  1-2\frac{mr}{\left(  r^{2}+a^{2}x^{2}\right)  }\right]
dt^{2}-4\frac{amr\left(  1-x^{2}\right)  }{\left(  r^{2}+a^{2}x^{2}\right)
}dtd\phi-\frac{r^{2}+a^{2}x^{2}}{r^{2}-2mr+a^{2}}dr^{2}\nonumber\\
&  -\frac{r^{2}+a^{2}x^{2}}{1-x^{2}}dx-\left[  \left(  r^{2}+a^{2}\right)
\left(  1-x^{2}\right)  +2\frac{mr}{\left(  r^{2}+a^{2}x^{2}\right)  }%
a^{2}\left(  1-x^{2}\right)  ^{2}\right]  d\phi^{2}. \label{xkerr-0}%
\end{align}

The zero-mass Kerr metric gives a flat metric:
\begin{equation}
ds^{2}=dt^{2}-\frac{r^{2}+a^{2}x^{2}}{r^{2}+a^{2}}dr^{2}-\frac{r^{2}%
+a^{2}x^{2}}{1-x^{2}}dx^{2}-\left(  r^{2}+a^{2}\right)  \left(  1-x^{2}%
\right)  d\varphi^{2}. \label{zzbpz-0}%
\end{equation}

Supposing the quantum corrected spacetime is axisymmetric, we take the line
element of quantum corrected spacetime as%
\begin{align}
ds^{2}  &  =\left[  1+a\left(  r,x\right)  \right]  dt^{2}-\left[  \frac
{r^{2}+a^{2}x^{2}}{r^{2}+a^{2}}-b\left(  r,x\right)  \right]  dr^{2}-\left[
\frac{r^{2}+a^{2}x^{2}}{1-x^{2}}-c\left(  r,x\right)  \right]  dx^{2}%
\nonumber\\
&  -\left[  \left(  r^{2}+a^{2}\right)  \left(  1-x^{2}\right)  -d\left(
r,x\right)  \right]  d\varphi^{2}+2\eta\left(  r,x\right)  dtd\varphi.
\label{zdczl-0}%
\end{align}

Substituting Eq. (\ref{zdczl-0}) into Eq. (\ref{zlfc-2-4}), we have
\begin{align}
a\left(  r,x\right)   &  =-\frac{2rm}{r^{2}+a^{2}x^{2}},\nonumber\\
b\left(  r,x\right)   &  =-\frac{2r\left(  r^{2}+a^{2}x^{2}\right)  m}{\left(
r^{2}+a^{2}\right)  ^{2}},\nonumber\\
c\left(  r,x\right)   &  =0,\nonumber\\
d\left(  r,x\right)   &  =-\frac{2a^{2}r\left(  x^{2}-1\right)  ^{2}m}%
{r^{2}+a^{2}x^{2}},\nonumber\\
\eta\left(  r,x\right)   &  =-\frac{2ar\left(  x^{2}-1\right)  m}{r^{2}%
+a^{2}x^{2}}. \label{zdczl-03-0}%
\end{align}

The quantum corrected result then reads
\begin{align}
ds^{2}  &  =\left(  1-\frac{2rm}{r^{2}+a^{2}x^{2}}\right)  dt^{2}-\left[
\frac{r^{2}+a^{2}x^{2}}{r^{2}+a^{2}}+\frac{2r\left(  r^{2}+a^{2}x^{2}\right)
m}{\left(  r^{2}+a^{2}\right)  ^{2}}\right]  dr^{2}-\frac{r^{2}+a^{2}x^{2}%
}{1-x^{2}}dx^{2}\nonumber\\
&  -\left[  \left(  r^{2}+a^{2}\right)  \left(  1-x^{2}\right)  +\frac
{2a^{2}r\left(  x^{2}-1\right)  ^{2}m}{r^{2}+a^{2}x^{2}}\right]  d\varphi
^{2}-2\frac{2ar\left(  x^{2}-1\right)  m}{r^{2}+a^{2}x^{2}}dtd\varphi.
\label{zdczl-2}%
\end{align}

The singularity of the metric appears at
\begin{equation}
x=\pm1.
\end{equation}
The singularity of the Ricci tensor appears at
\begin{align}
x  &  =\pm1,\nonumber\\
r  &  =m\pm\sqrt{m^{2}-a^{2}},\nonumber\\
r  &  =-m\pm\sqrt{m^{2}-a^{2}}.
\end{align}
The singularities of the Riemann tensor, the Weyl tensor, and the Ricci scalar
appear at%
\begin{align}
r  &  =m\pm\sqrt{m^{2}-a^{2}},\nonumber\\
r  &  =-m\pm\sqrt{m^{2}-a^{2}}.
\end{align}

The spacetime with the metric (\ref{zdczl-2}) has an event horizon at $r=\pm
a$ and an infinite redshift surface at $r=m\pm\sqrt{m^{2}-ax^{2}}=m\pm
\sqrt{m^{2}-a\cos^{2}\theta}$.

Comparing with the Kerr case, we can see that quantum corrected metric
(\ref{zdczl-2}) is a small-mass Kerr metric, i.e., the leading term of the
expansion of the Kerr metric around $m=0$.

\subsection{Cylindrically symmetric quantum fluctuation}

In this section, we consider a flat spacetime with an cylindrically symmetric
quantum fluctuation.

We write the metric of a flat space as a zero-mass Curzon spacetime who is a
flat spacetime with zero curvatures.

The $1+3$-dimensional Curzon metric is \cite{griffiths2009exact}%
\begin{equation}
ds^{2}=\exp\left(  -\frac{2m}{\sqrt{\rho^{2}+z^{2}}}\right)  dt^{2}%
-\exp\left(  \frac{2m}{\sqrt{\rho^{2}+z^{2}}}\right)  \left[  \exp\left(
-\frac{m^{2}\rho^{2}}{\left(  \rho^{2}+z^{2}\right)  ^{2}}\right)  \left(
d\rho^{2}+dz^{2}\right)  +\rho^{2}d\varphi^{2}\right]  .
\end{equation}
By the coordinate transformation $\rho=\left(  r^{2}-2mr\right)  ^{1/2}%
\sin\theta$ and $z=\left(  r-m\right)  \cos\theta$, we arrive at the Curzon
metric represented by the spherical coordinate $\left(  t,r,\theta
,\varphi\right)  $:%
\begin{align}
ds^{2}  &  =\exp\left(  -\frac{2m}{\sqrt{r^{2}-2mr+m^{2}\cos^{2}\theta}%
}\right)  dt^{2}-\exp\left(  \frac{2m}{\sqrt{r^{2}-2mr+m^{2}\cos^{2}\theta}%
}\right) \nonumber\\
&  \times\left[  \exp\left(  -\frac{m^{2}\left(  r^{2}-2mr\right)  \sin
^{2}\theta}{\left(  r^{2}-2mr+m^{2}\cos^{2}\theta\right)  ^{2}}\right)
\left(  r^{2}-2mr+m^{2}\sin^{2}\theta\right)  \left(  \frac{1}{r^{2}%
-2mr}dr^{2}+d\theta^{2}\right)  \right. \nonumber\\
&  \left.  +\left(  r^{2}-2mr\right)  \sin^{2}\theta d\varphi^{2}\right]  .
\label{curzon-1}%
\end{align}

Next, by the coordinate transformation $\cos\theta=x$, we arrive at the Curzon
metric represented by the coordinate $\left(  t,r,x,\varphi\right)  $:%
\begin{align}
ds^{2}  &  =\exp\left(  -\frac{2m}{\sqrt{r^{2}-2mr+m^{2}x^{2}}}\right)
dt^{2}-\exp\left(  \frac{2m}{\sqrt{r^{2}-2mr+m^{2}x^{2}}}\right) \nonumber\\
&  \times\left\{  \exp\left(  -\frac{m^{2}\left(  r^{2}-2mr\right)  \left(
1-x^{2}\right)  }{\left(  r^{2}-2mr+m^{2}x^{2}\right)  ^{2}}\right)  \left[
r^{2}-2mr+m^{2}\left(  1-x^{2}\right)  \right]  \right. \label{curzon-2}\\
&  \times\left(  \frac{1}{r^{2}-2mr}dr^{2}+\frac{1}{1-x^{2}}dx^{2}\right)
\left.  +\left(  r^{2}-2mr\right)  \left(  1-x^{2}\right)  d\varphi
^{2}\right\}  .
\end{align}

The zero-mass Curzon metric gives a flat metric:%
\begin{equation}
ds^{2}=dt^{2}-dr^{2}-\frac{r^{2}}{1-x^{2}}dx^{2}-r^{2}\left(  1-x^{2}\right)
d\varphi^{2}. \label{curzon-m-0}%
\end{equation}

Similar procedure gives%
\begin{equation}
ds^{2}=\left(  1-\frac{C}{r}\right)  dt^{2}-\left(  1+\frac{C}{r}\right)
dr^{2}-\frac{r^{2}}{1-x^{2}}dx^{2}-r^{2}\left(  1-x^{2}\right)  d\varphi^{2}.
\label{curzon-m-C-0}%
\end{equation}

The quantum fluctuation is small. To determine the constant $C$, we compare
Eq. (\ref{curzon-m-C-0}) with the Curzon metric with a small mass. The
comparison gives $C=2m$. The quantum corrected metric is%
\begin{equation}
ds^{2}=\left(  1-\frac{2m}{r}\right)  dt^{2}-\left(  1+\frac{2m}{r}\right)
dr^{2}-\frac{r^{2}}{1-x^{2}}dx^{2}-r^{2}\left(  1-x^{2}\right)  d\varphi^{2}.
\label{curzon-m-01}%
\end{equation}

This is the Curzon metric with a small mass. The singularity of the
metric\ (\ref{curzon-m-01}) is at $r=0$ and $x=\pm1$. The singularity of the
Ricci tensor, the Riemann tensor, the Weyl tensor, and the Ricci scalar is at
$r=0$ and $r=2m$. The infinite redshift surface is at $r=2m$. This spacetime
has no event horizon.

\section{Quantum fluctuation in Schwarzschild spacetime \label{QFS}}

The classical Schwarzschild spacetime%
\begin{equation}
ds^{2}=\left(  1-\frac{2m}{r}\right)  dt^{2}-\frac{1}{1-\frac{2m}{r}}%
dr^{2}-r^{2}d\theta^{2}-r^{2}\sin^{2}\theta d\varphi^{2} \label{jdswx-0-0}%
\end{equation}
is a bound state of gravitational fields. Suppose there is a quantum
fluctuation in the form%
\begin{equation}
ds^{2}=\left(  1-\frac{2m}{r}\right)  dt^{2}-\frac{1}{1-\frac{2m}{r}}%
dr^{2}-r^{2}d\theta^{2}-r^{2}\sin^{2}\theta d\varphi^{2}+2\sin^{2}\theta
f\left(  r\right)  dtd\varphi\label{jtqzl-0}%
\end{equation}

Substituting Eqs. (\ref{jdswx-0-0}) and (\ref{jtqzl-0}) into the equation of
quantum fluctuations, Eq. (\ref{zlfc-2-4}), we obtain an equation for
$f\left(  r\right)  $:
\begin{equation}
\left(  2r-3m\right)  f\left(  r\right)  -r^{2}\left(  r-2m\right)
f^{\prime\prime}\left(  r\right)  =0. \label{swxzleq-0-0}%
\end{equation}
The solution is
\begin{equation}
f\left(  r\right)  =Cf_{1}\left(  r\right)  +C^{\prime}f_{2}\left(  r\right)
\label{swxzlj-0-0}%
\end{equation}
with
\begin{align}
f_{1}\left(  r\right)   &  =r^{\frac{1-\sqrt{7}}{2}}\left(  r-2m\right)
\text{ }_{2}\operatorname{F}_{1}\left(  \frac{5-\sqrt{7}}{2},\frac{-1-\sqrt
{7}}{2},1-\sqrt{7},\frac{r}{2m}\right)  ,\label{fxdl-0}\\
f_{2}\left(  r\right)   &  =r^{\frac{1+\sqrt{7}}{2}}\left(  r-2m\right)
\text{ }_{2}\operatorname{F}_{1}\left(  \frac{5+\sqrt{7}}{2},\frac{-1+\sqrt
{7}}{2},1+\sqrt{7},\frac{r}{2m}\right)  . \label{fxdl-1}%
\end{align}

Next we determine the constants $C$ and $C^{\prime}$.

For $0<r\leq2m$, $f_{1}\left(  r\right)  $ and $f_{2}\left(  r\right)  $ are
both real functions. This requires that $C$ and $C^{\prime}$ should be real numbers.

For $r>2m$, $f_{1}\left(  r\right)  $ and $f_{2}\left(  r\right)  $ are both
complex functions. We rewrite Eq. (\ref{swxzlj-0-0}) as%
\begin{equation}
f\left(  r\right)  =\left(  C\operatorname{Re}f_{1}+C^{\prime}%
\operatorname{Re}f_{2}\right)  +i\left(  C\operatorname{Im}f_{1}+C^{\prime
}\operatorname{Im}f_{2}\right)  \label{f-sxb-0}%
\end{equation}
Requiring $f\left(  r\right)  $ is a real function, we must have
$C\operatorname{Im}f_{1}+C^{\prime}\operatorname{Im}f_{2}=0$, i.e.,%
\begin{equation}
-\frac{C^{\prime}}{C}=\frac{\operatorname{Im}f_{1}}{\operatorname{Im}f_{2}%
}\equiv k. \label{swxzl-xsgx-00}%
\end{equation}

To determine the constant $k=-C^{\prime}/C$, we expand $f_{1}\left(  r\right)
$ and $f_{2}\left(  r\right)  $ at $r\rightarrow\infty$:%
\begin{equation}
k=-\frac{1}{m^{\sqrt{7}}}\frac{\left(  4+\sqrt{7}\right)  \Gamma\left(
1-\frac{\sqrt{7}}{2}\right)  \Gamma\left(  \frac{5}{2}+\frac{\sqrt{7}}%
{2}\right)  }{3\times8^{\sqrt{7}}\Gamma\left(  1+\frac{\sqrt{7}}{2}\right)
\Gamma\left(  \frac{5}{2}-\frac{\sqrt{7}}{2}\right)  }.
\end{equation}
Then we have $f\left(  r\right)  =C\left(  \operatorname{Re}f_{1}%
-k\operatorname{Re}f_{2}\right)  $. That is, $f\left(  r\right)  $ is
real\ for $r>2m$, for $C$ and $k$ are real.

Now we rewrite $f\left(  r\right)  $, by substituting Eq. (\ref{swxzl-xsgx-00}%
) into Eq. (\ref{swxzlj-0-0}), as%
\begin{equation}
f\left(  r\right)  =C\left[  f_{1}\left(  r\right)  -kf_{2}\left(  r\right)
\right]  . \label{fr-k-00}%
\end{equation}

The quantum corrected result then reads%
\begin{equation}
ds^{2}=\left(  1-\frac{2m}{r}\right)  dt^{2}-\frac{1}{1-\frac{2m}{r}}%
dr^{2}-r^{2}d\theta^{2}-r^{2}\sin^{2}\theta d\varphi^{2}+2C\sin^{2}%
\theta\left[  f_{1}\left(  r\right)  -kf_{2}\left(  r\right)  \right]
dtd\varphi.
\end{equation}

The singularities of the metric and various curvatures, such as the Riemann
tensor, are at $r=0$ and $r=2m$, as that in the classical Schwarzschild
spacetime. This spacetime has an event horizon and infinite redshift surface
at $r=2m$.

\section{Conclusion \label{Conclusion}}

In this paper, we construct the quantum effective action for gravitational
fields by the effective potential method in quantum field theory. Based on the
quantum effective action, we construct the equation for the leading
contribution of the quantum fluctuation. The equation of quantum fluctuation
is a linear equation. If the higher-order contribution of quantum fluctuations
is taken into account, the equation for the quantum fluctuation will become a
nonlinear equation.

The premise of the effective potential method is that one must first have a
classical solution of the field equation, and then calculate the quantum
fluctuation for each classical solution. For matter fields, the solution of
the classical field equation serves as an effective potential in the equation
of quantum fluctuation. For gravitational fields, the solution of the
classical gravitational field equation, the Einstein equation, also serves as
a effective potential but in a more complex form. Different classical
solutions give different effective potentials. The advantage of this method is
that we can deal with the quantum effect by quantum field theory method even
for bound states.

Flat spacetime is a classical vacuum. In this paper, we calculate the quantum
fluctuation in flat spacetime. When there occurs a quantum fluctuation, a
spacetime is created. The spacetime created from quantum fluctuations,
although the mass is very small, may be of the magnitude of the Planck mass,
is a seed of spacetime, namely, a baby spacetime. Some of the baby spacetime
will grow into our present spacetime. With the growth of spacetime, the
quantum effect will become less and less obvious, closer and closer to the
classical spacetime described by the Einstein equation.

We consider the quantum fluctuations in the Schwarzschild spacetime. The
Schwarzschild spacetime is a bound state of gravitational field. The usual
quantum-field-theory treatment based on the scattering-state perturbation
cannot deal with bound states. We deal with this problem using the effective
potential method.

In principle, quantum fluctuations occur randomly and arbitrarily. However,
for ease of calculation, only some quantum fluctuations with special
symmetries are considered. Quantum fluctuations are of course not limited to
special symmetry cases. Like finding the exact solution for the Einstein
equation, we can only treat such relatively simple cases.

In this paper, we construct an equation for quantum fluctuation by solving the
quantum effective action. The quantum effective action is a spectral function
\cite{vassilevich2003heat,dai2009number,dai2010approach}. In Ref.
\cite{dai2010approach}, we construct an equation for one-loop effective
action. In future works, we may consider the quantum effect in such a way.


\acknowledgments

We are very indebted to Dr G. Zeitrauman for his encouragement. This work is supported in part by Special Funds for theoretical physics Research Program of the NSFC under Grant No. 11947124, and NSFC under Grant Nos. 11575125 and 11675119.




\begin{thebibliography}{10}

\bibitem{huang1992quarks}
K.~Huang, {\em Quarks, leptons \& gauge fields}.
\newblock World Scientific, 1992.

\bibitem{weinberg2005quantum}
S.~Weinberg, {\em The Quantum Theory of Fields: Volume 2, Modern Applications}.
\newblock Cambridge University Press, 2005.

\bibitem{graham2003negative}
N.~Graham and K.~D. Olum, {\it Negative energy densities in quantum field
  theory with a background potential},  {\em Physical Review D} {\bf 67}
  (2003), no.~8 085014.

\bibitem{graham2003casimir}
N.~Graham, R.~Jaffe, V.~Khemani, M.~Quandt, M.~Scandurra, and H.~Weigel, {\it
  Casimir energies in light of quantum field theory},  {\em Physics Letters B}
  {\bf 572} (2003), no.~3 196--201.

\bibitem{graham2002calculating}
N.~Graham, R.~Jaffe, V.~Khemani, M.~Quandt, M.~Scandurra, and H.~Weigel, {\it
  Calculating vacuum energies in renormalizable quantum field theories: A new
  approach to the Casimir problem},  {\em Nuclear Physics B} {\bf 645} (2002),
  no.~1 49--84.

\bibitem{graham1999energy}
N.~Graham and R.~Jaffe, {\it Energy, central charge, and the BPS bound for 1+
  1-dimensional supersymmetric solitons},  {\em Nuclear Physics B} {\bf 544}
  (1999), no.~1 432--447.

\bibitem{graham2009spectral}
N.~Graham, M.~Quandt, and H.~Weigel, {\em Spectral methods in quantum field
  theory}, vol.~777.
\newblock Springer, 2009.

\bibitem{rahi2009scattering}
S.~J. Rahi, T.~Emig, N.~Graham, R.~L. Jaffe, and M.~Kardar, {\it Scattering
  theory approach to electrodynamic Casimir forces},  {\em Physical Review D}
  {\bf 80} (2009), no.~8 085021.

\bibitem{weigel2018spectral}
H.~Weigel, M.~Quandt, and N.~Graham, {\it Spectral methods for coupled channels
  with a mass gap},  {\em Physical Review D} {\bf 97} (2018), no.~3 036017.

\bibitem{li2021duality}
W.-D. Li and W.-S. Dai, {\it Duality family of scalar field},  {\em Nuclear
  Physics B} {\bf 972} (2021) 115569.

\bibitem{griffiths2009exact}
J.~B. Griffiths and J.~Podolsk{\`y}, {\em Exact space-times in Einstein's
  general relativity}.
\newblock Cambridge University Press, 2009.

\bibitem{stephani2009exact}
H.~Stephani, D.~Kramer, M.~MacCallum, C.~Hoenselaers, and E.~Herlt, {\em Exact
  Solutions of Einstein's Field Equations}.
\newblock Cambridge Monographs on Mathematical Physics. Cambridge University
  Press, 2009.

\bibitem{dai2000electromagnetic}
W.-S. Dai, X.-H. Guo, H.-Y. Jin, and X.-Q. Li, {\it Electromagnetic radiation
  of baryons containing two heavy quarks},  {\em Physical Review D} {\bf 62}
  (2000), no.~11 114026.

\bibitem{woodard2009far}
R.~P. Woodard, {\it How far are we from the quantum theory of gravity?},  {\em
  Reports on Progress in Physics} {\bf 72} (2009), no.~12 126002.

\bibitem{rovelli2011loop}
C.~Rovelli, {\it Loop quantum gravity: the first 25 years},  {\em Classical and
  Quantum Gravity} {\bf 28} (2011), no.~15 153002.

\bibitem{hollands2015quantum}
S.~Hollands and R.~M. Wald, {\it Quantum fields in curved spacetime},  {\em
  Physics Reports} {\bf 574} (2015) 1--35.

\bibitem{draper2020finite}
T.~Draper, B.~Knorr, C.~Ripken, and F.~Saueressig, {\it Finite Quantum Gravity
  Amplitudes: No Strings Attached},  {\em Physical Review Letters} {\bf 125}
  (2020), no.~18 181301.

\bibitem{geng2020distance}
H.~Geng, {\it Distance conjecture and de-Sitter quantum gravity},  {\em Physics
  Letters B} {\bf 803} (2020) 135327.

\bibitem{bergshoeff2018nonrelativistic}
E.~Bergshoeff, J.~Gomis, and Z.~Yan, {\it Nonrelativistic string theory and
  T-duality},  {\em Journal of High Energy Physics} {\bf 2018} (2018), no.~11
  1--23.

\bibitem{kutasov2015constraining}
D.~Kutasov, T.~Maxfield, I.~Melnikov, and S.~Sethi, {\it Constraining de Sitter
  space in string theory},  {\em Physical review letters} {\bf 115} (2015),
  no.~7 071305.

\bibitem{eberhardt2019string}
L.~Eberhardt and M.~R. Gaberdiel, {\it String theory on AdS3 and the symmetric
  orbifold of Liouville theory},  {\em Nuclear Physics B} {\bf 948} (2019)
  114774.

\bibitem{cano2018alpha}
P.~A. Cano, P.~Meessen, T.~Ort{\'\i}n, and P.~F. Ramirez, {\it
  $\alpha^\prime$-corrected black holes in String Theory},  {\em Journal of
  High Energy Physics} {\bf 2018} (2018), no.~5 110.

\bibitem{pioline2019string}
B.~Pioline, {\it String theory integrands and supergravity divergences},  {\em
  Journal of High Energy Physics} {\bf 2019} (2019), no.~2 1--35.

\bibitem{antoniadis2018effective}
I.~Antoniadis, A.~Delgado, C.~Markou, and S.~Pokorski, {\it The effective
  supergravity of little string theory},  {\em The European Physical Journal C}
  {\bf 78} (2018), no.~2 1--9.

\bibitem{cribiori2020sitter}
N.~Cribiori, R.~Kallosh, A.~Linde, and C.~Roupec, {\it de Sitter Minima from
  M-theory and String theory},  {\em Physical Review D} {\bf 101} (2020), no.~4
  046018.

\bibitem{hull2020black}
C.~Hull, E.~Marcus, K.~Stemerdink, and S.~Vandoren, {\it Black holes in string
  theory with duality twists},  {\em Journal of High Energy Physics} {\bf 2020}
  (2020), no.~7 1--58.

\bibitem{heslop2018m}
P.~Heslop and A.~E. Lipstein, {\it M-theory beyond the supergravity
  approximation},  {\em Journal of High Energy Physics} {\bf 2018} (2018),
  no.~2 1--17.

\bibitem{ashtekar1998quantum}
A.~Ashtekar, J.~Baez, A.~Corichi, and K.~Krasnov, {\it Quantum geometry and
  black hole entropy},  {\em Physical Review Letters} {\bf 80} (1998), no.~5
  904.

\bibitem{ashtekar1999isolated}
A.~Ashtekar, A.~Corichi, and K.~Krasnov, {\it Isolated horizons: the classical
  phase space},  {\em arXiv preprint gr-qc/9905089} (1999).

\bibitem{ashtekar2000quantum}
A.~Ashtekar, J.~Baez, and K.~Krasnov, {\it Quantum geometry of isolated
  horizons and black hole entropy},  {\em arXiv preprint gr-qc/0005126} (2000).

\bibitem{ashtekar2000isolated}
A.~Ashtekar, S.~Fairhurst, and B.~Krishnan, {\it Isolated horizons: Hamiltonian
  evolution and the first law},  {\em Physical Review D} {\bf 62} (2000),
  no.~10 104025.

\bibitem{ashtekar2001mechanics}
A.~Ashtekar, C.~Beetle, and J.~Lewandowski, {\it Mechanics of rotating isolated
  horizons},  {\em Physical Review D} {\bf 64} (2001), no.~4 044016.

\bibitem{ashtekar2002geometry}
A.~Ashtekar, C.~Beetle, and J.~Lewandowski, {\it Geometry of generic isolated
  horizons},  {\em Classical and Quantum Gravity} {\bf 19} (2002), no.~6 1195.

\bibitem{kelly2020black}
J.~G. Kelly, R.~Santacruz, and E.~Wilson-Ewing, {\it Black hole collapse and
  bounce in effective loop quantum gravity},  {\em Classical and Quantum
  Gravity} {\bf 38} (2020), no.~4 04LT01.

\bibitem{yang2019quantum}
Z.~Yang, {\it The quantum gravity dynamics of near extremal black holes},  {\em
  Journal of High Energy Physics} {\bf 2019} (2019), no.~5 1--36.

\bibitem{alesci2019quantum}
E.~Alesci, S.~Bahrami, and D.~Pranzetti, {\it Quantum gravity predictions for
  black hole interior geometry},  {\em Physics Letters B} {\bf 797} (2019)
  134908.

\bibitem{li2018scalar}
W.-D. Li, Y.-Z. Chen, and W.-S. Dai, {\it Scalar scattering in Schwarzschild
  spacetime: Integral equation method},  {\em Physics Letters B} (2018).

\bibitem{li2019scattering}
W.-D. Li, Y.-Z. Chen, and W.-S. Dai, {\it Scattering state and bound state of
  scalar field in Schwarzschild spacetime: Exact solution},  {\em Annals of
  Physics} {\bf 409} (2019) 167919.

\bibitem{li2021scalar}
S.-L. Li, Y.-Y. Liu, W.-D. Li, and W.-S. Dai, {\it Scalar field in
  Reissner--Nordstr{\"o}m spacetime: Bound state and scattering state (with
  appendix on eliminating oscillation in partial sum approximation of periodic
  function)},  {\em Annals of Physics} {\bf 432} (2021) 168578.

\bibitem{casadio2020bootstrapped}
R.~Casadio and I.~Kuntz, {\it Bootstrapped Newtonian quantum gravity},  {\em
  The European Physical Journal C} {\bf 80} (2020), no.~6 1--7.

\bibitem{maxfield2019quantum}
H.~Maxfield, {\it Quantum corrections to the BTZ black hole extremality bound
  from the conformal bootstrap},  {\em Journal of High Energy Physics} {\bf
  2019} (2019), no.~12 1--44.

\bibitem{bautista2019quantum}
T.~Bautista, A.~Dabholkar, and H.~Erbin, {\it Quantum gravity from timelike
  Liouville theory},  {\em Journal of High Energy Physics} {\bf 2019} (2019),
  no.~10 1--41.

\bibitem{afkhami2019fast}
N.~Afkhami-Jeddi, T.~Hartman, and A.~Tajdini, {\it Fast conformal bootstrap and
  constraints on 3d gravity},  {\em Journal of High Energy Physics} {\bf 2019}
  (2019), no.~5 1--25.

\bibitem{casadio2020quantum}
R.~Casadio, M.~Lenzi, and A.~Ciarfella, {\it Quantum black holes in
  bootstrapped Newtonian gravity},  {\em Physical Review D} {\bf 101} (2020),
  no.~12 124032.

\bibitem{falls2019aspects}
K.~G. Falls, D.~F. Litim, and J.~Schr{\"o}der, {\it Aspects of asymptotic
  safety for quantum gravity},  {\em Physical Review D} {\bf 99} (2019), no.~12
  126015.

\bibitem{aprile2018quantum}
F.~Aprile, J.~Drummond, P.~Heslop, and H.~Paul, {\it Quantum gravity from
  conformal field theory},  {\em Journal of High Energy Physics} {\bf 2018}
  (2018), no.~1 1--26.

\bibitem{oriti2018black}
D.~Oriti, D.~Pranzetti, and L.~Sindoni, {\it Black holes as quantum gravity
  condensates},  {\em Physical Review D} {\bf 97} (2018), no.~6 066017.

\bibitem{hennigar2020lower}
R.~A. Hennigar, D.~Kubiz{\v{n}}{\'a}k, R.~B. Mann, and C.~Pollack, {\it
  Lower-dimensional Gauss--Bonnet gravity and BTZ black holes},  {\em Physics
  Letters B} {\bf 808} (2020) 135657.

\bibitem{cotler2020low}
J.~Cotler, K.~Jensen, and A.~Maloney, {\it Low-dimensional de Sitter quantum
  gravity},  {\em Journal of High Energy Physics} {\bf 2020} (2020),
  no.~1905.03780 1--103.

\bibitem{amendola2017quantum}
L.~Amendola, N.~Burzill{\`a}, and H.~Nersisyan, {\it Quantum gravity inspired
  nonlocal gravity model},  {\em Physical Review D} {\bf 96} (2017), no.~8
  084031.

\bibitem{ruzziconi2021conservation}
R.~Ruzziconi and C.~Zwikel, {\it Conservation and integrability in
  lower-dimensional gravity},  {\em Journal of High Energy Physics} {\bf 2021}
  (2021), no.~4 1--39.

\bibitem{ma2020bounds}
L.~Ma and H.~L{\"u}, {\it Bounds on photon spheres and shadows of charged black
  holes in Einstein-Gauss-Bonnet-Maxwell gravity},  {\em Physics Letters B}
  {\bf 807} (2020) 135535.

\bibitem{ma2021d}
L.~Ma, Y.-Z. Li, and H.~L{\"u}, {\it D= 5 rotating black holes in
  Einstein-Gauss-Bonnet gravity: mass and angular momentum in extremality},
  {\em Journal of High Energy Physics} {\bf 2021} (2021), no.~1 1--28.

\bibitem{ma2020vacua}
L.~Ma and H.~L{\"u}, {\it Vacua and exact solutions in lower-D limits of EGB},
  {\em The European Physical Journal C} {\bf 80} (2020), no.~12 1--10.

\bibitem{perez2019dark}
A.~Perez and D.~Sudarsky, {\it Dark energy from quantum gravity discreteness},
  {\em Physical review letters} {\bf 122} (2019), no.~22 221302.

\bibitem{gielen2018cosmological}
S.~Gielen and D.~Oriti, {\it Cosmological perturbations from full quantum
  gravity},  {\em Physical Review D} {\bf 98} (2018), no.~10 106019.

\bibitem{Chen2021model}
C.~Yu-Zhu, C.~Yu-Jie, L.~Shi-Lin, Z.~Fu-Lin, and D.~Wu-Sheng, {\it Model of
  black hole and white hole in Minkowski spacetime},  {\em The European
  Physical Journal C} {\bf 81} (2021), no.~12 1--10.

\bibitem{chen2018entropy}
Y.-Z. Chen, W.-D. Li, and W.-S. Dai, {\it Why the entropy of spacetime is
  independent of species of particles: the species problem},  {\em The European
  Physical Journal C} {\bf 78} (2018), no.~8 635.

\bibitem{weinberg1973gravitation}
S.~Weinberg and R.~V. Wagoner, {\it Gravitation and cosmology: principles and
  applications of the general theory of relativity},  {\em Physics Today} {\bf
  26} (1973), no.~6 57.

\bibitem{burns1991modern}
R.~Burns, B.~Dubrovin, A.~Fomenko, and S.~Novikov, {\em Modern Geometry -
  Methods and Applications: Part I: The Geometry of Surfaces, Transformation
  Groups, and Fields}.
\newblock Springer New York, 1991.

\bibitem{economou2013green}
E.~Economou, {\em Green's Functions in Quantum Physics}.
\newblock Springer Series in Solid-State Sciences. Springer Berlin Heidelberg,
  2013.

\bibitem{ryder1996quantum}
L.~Ryder and H.~Ryder, {\em Quantum Field Theory}.
\newblock Cambridge University Press, 1996.

\bibitem{ohanian2013gravitation}
H.~Ohanian and R.~Ruffini, {\em Gravitation and Spacetime}.
\newblock Cambridge University Press, 2013.

\bibitem{vassilevich2003heat}
D.~V. Vassilevich, {\it Heat kernel expansion: user's manual},  {\em Physics
  reports} {\bf 388} (2003), no.~5 279--360.

\bibitem{dai2009number}
W.-S. Dai and M.~Xie, {\it The number of eigenstates: counting function and
  heat kernel},  {\em Journal of High Energy Physics} {\bf 2009} (2009), no.~02
  033.

\bibitem{dai2010approach}
W.-S. Dai and M.~Xie, {\it An approach for the calculation of one-loop
  effective actions, vacuum energies, and spectral counting functions},  {\em
  Journal of High Energy Physics} {\bf 2010} (2010), no.~6 1--29.

\end{thebibliography}



\end{document}